\documentclass[aps,prb,twocolumn,superscriptaddress,showpacs]{revtex4-1}

\usepackage{bm}					
\usepackage{graphicx}
\usepackage{amsmath,amsfonts,amssymb}	
\usepackage{latexsym} 				
\usepackage{xcolor}
\usepackage{subfigure}

\begin{document}

\title{Spectral Properties and Local Density of States\\
of Disordered Quantum Hall Systems with Rashba Spin-Orbit Coupling}

\author{Daniel \surname{Hernang\'{o}mez-P\'{e}rez}}

\affiliation{Laboratoire de Physique et Mod\'elisation des Milieux Condens\'es, CNRS and 
Universit\'e Joseph Fourier, B.P. 166, 25 rue des Martyrs, F-38042 Grenoble, France}

\author{Jascha Ulrich}
\affiliation{Institute of Physics and JARA-FIT, RWTH Aachen University,
D-52074 Aachen, Germany}

\affiliation{Institut N\'eel, CNRS and Universit\'e Joseph Fourier, B.P. 166, 25 rue des Martyrs, 
F-38042 Grenoble, France}

\author{Serge Florens}
\affiliation{Institut N\'eel, CNRS and Universit\'e Joseph Fourier, B.P. 166, 25 rue des Martyrs, 
F-38042 Grenoble, France}

\author{Thierry Champel}

\affiliation{Laboratoire de Physique et Mod\'elisation des Milieux Condens\'es, CNRS and 
Universit\'e Joseph Fourier, B.P. 166, 25 rue des Martyrs, F-38042 Grenoble, France}

\date{\today}


\begin{abstract}
We theoretically investigate the spectral properties and the spatial dependence
of the local density of states (LDoS) in disordered two-dimensional electron gases
(2DEG) in the quantum Hall regime, taking into account the combined presence of 
electrostatic disorder, random Rashba spin-orbit interaction, and finite Zeeman 
coupling. To this purpose, we extend a coherent-state Green's function formalism 
previously proposed for spinless 2DEG in the presence of smooth arbitrary disorder, 
that here incorporates the nontrivial coupling between the orbital and spin
degrees of freedom into the electronic drift states. The formalism
allows us to obtain analytical and controlled nonperturbative expressions of
the energy spectrum in arbitrary locally flat disorder potentials with both random
electric fields and Rashba coupling. As an illustration of this theory, we
derive analytical microscopic expressions for the LDoS in different
temperature regimes which can be used as a starting point to interpret scanning
tunneling spectroscopy data at high magnetic fields. In this context, we study the
spatial dependence and linewidth of the LDoS peaks and explain an
experimentally-noticed correlation between the spatial dispersion of the spin-orbit
splitting and the local extrema of the potential landscape.
\end{abstract}

\maketitle
\section{Introduction}\label{Introduction}
\subsection{Motivation}

The study of spin-orbit (SO) induced phenomena in semiconductor heterostructures has evolved during the last two decades into a rich research subfield of spintronics both due to the interesting fundamental physics involved\cite{Zutic2004} and the potential applications, which span from information processing devices to quantum computation\cite{Awschalon2013}. One of the most important goals in this area consists in the local injection, transfer, manipulation and detection of spin in a controllable and coherent way, and it has been recognized that the SO coupling is a particularly well-adapted tool. Moreover, one expects to be able to control the spin degree of freedom using electric fields created by local voltage gates since the charge and spin degrees of freedom become coupled. The situation could allow an implementation of some spintronic devices in disordered two-dimensional electron gases (2DEG) based in the spin-field effect transistor\cite{Datta1990} or its counterpart in the quantum Hall regime\cite{Giovannetti2012, Komiyama2012} where one takes advantage of the existence of spin-resolved quantum Hall edge channels.

In 2DEG at the interface of III-V semiconductors with zincblende crystal
structure  there exists an
intrinsic solid-state SO coupling.
We can distinguish two main
contributions at lowest order in the momentum:
Rashba\cite{Rashba1960,Rashba1984} and Dresselhaus\cite{Dresselhaus1955},
characterized respectively by SO coupling parameters $\alpha$ and
$\beta$ with clear and distinct physical origins. 
The Rashba coupling arises
as a consequence of the lack of structure inversion symmetry in the confining
potential, while the Dresselhaus coupling takes its origin in bulk inversion asymmetry
and therefore just depends on the crystal lattice
structure\cite{Schliemann2006}. As a consequence, the Rashba coupling parameter
$\alpha$ is proportional to the gradient of the potential in the
perpendicular direction (being tunable by external voltage
gates\cite{Enoki1997}), while the Dresselhaus
parameter  $\beta$ is only sensitive to deep changes in the crystal-lattice affecting the
structural integrity of the heterostructure. 
Both Rashba and Dresselhaus contributions may be present
in a given semiconductor heterostructure and which of them is dominant depends
on material parameters and the perpendicular potential
gradients\cite{DasSarma2003}. For instance, one usually has pure Dresselhaus coupling in GaAs, or pure Rashba coupling in InSb, while both can equally contribute  in InAs.

The particular situation affecting a specific heterostructure can be experimentally determined
 using weak localization to
antilocalization transitions\cite{Miller2003} or photocurrent measurements of
the angular distribution of the spin density\cite{Prettl2004}. The Rashba parameter $\alpha$ has been estimated  by an
analysis of the nodes of the beating patterns in the Shubnikov-de Haas
oscillations of the longitudinal magnetoresistance under magnetic fields in
InGaAs/InAlAs (Ref. \onlinecite{Enoki1997}) and HgTe\cite{Molenkamp2004}.
 Quite recently, it has also been determined by scanning
tunneling spectroscopy (STS) in InSb surface gases\cite{Morgenstern2010,Morgenstern2012} at high
magnetic fields, with an extraction of the coupling constant $\alpha$ from the positions of the nodes of
the density of states (DoS). 

Importantly, the STS technique gives primarily access to the {\em local} density of states
(LDoS), and thus provides an opportunity to reveal the spatial fluctuations of the SO coupling parameter on a local scale. Indeed,  random spatial fluctuations of the Rashba SO coupling are naturally expected in semiconductor heterostructures, because the electric field perpendicular to the well is created by dopant ions whose concentration unavoidably fluctuates spatially \cite{Sherman2003}. Therefore, the electron motion in the 2DEG is affected, in principle, by two different kinds of disorder, which are {\em a priori} locally uncorrelated: the electrostatic (in-plane) disorder potential $V({\bf r})$ acting on the charge motion, and a fluctuating SO Rashba coupling $\alpha({\bf r})$. These fluctuations in the Rashba coefficient
are known to influence the spin dynamics under weak but classical magnetic
fields by inducing memory effects for the spin
relaxation\cite{Sherman2005}. 

Recent STS measurements by Morgenstern {\em et al.} \cite{Morgenstern2010,Morgenstern2012}  in InSb surface gases at high magnetic fields (within the quantum Hall regime)
have shown that the energy spin splitting  indeed
varies spatially, what one could naively directly attribute to spatial fluctuations of the Rashba coupling. More precisely, a spin-split LDoS has been clearly observed only when the tip position is located close to the local extrema of the disorder potential, the energy spin splitting being typically\cite{Morgenstern2012} larger at hills of the potential landscape (close to local maxima), and smaller at valleys (near minima). In regions where the gradients of the potential landscape are strong, the energy spin splitting could not be determined due to the enlarged linewidth of the LDoS. 

In this paper we provide a simple explanation for this observed puzzling correlation\cite{Morgenstern2010,Morgenstern2012} between the spatial dispersion of the spin splitting and the disorder potential landscape.  We stress that these recent STS measurements  have been performed at high magnetic fields in the quantum Hall regime, an important aspect which has to be carefully taken into account within the theoretical interpretation of the LDoS characteristic features.

\subsection{Spectral properties  of 2DEG in high magnetic fields with Rashba spin-orbit interaction}

Numerous theoretical works have already considered the spectral properties of 2DEG with a uniform
Rashba SO coupling in strong quantizing magnetic fields. Under these conditions, it is required to include in the description the Zeeman coupling, which also contributes to the energy spin splitting. The resulting  energy levels in the absence of potential energy are known since several decades\cite{Rashba1960,Rashba1984}, with the result

\begin{equation}
  E_{n,\lambda}=\hbar \omega_{c}\left[n-\dfrac{\lambda}{2}\sqrt{(1-Z)^{2}+nS^{2}} \right]. \label{LLpure}
\end{equation}
Here
$n=0,1,2,\ldots $ is an integer and
  $\lambda=\pm1$ is the SO index which corresponds to two different
projections along the Rashba dependent spin axis (note that for $n=0$, only the projection $\lambda=-1$ is allowed). The corresponding eigenstates have a spinorial structure composed out of adjacent Landau level states which are associated with quantizations of the electronic cyclotron orbits in a magnetic field.  
 The dominant energy scale at high magnetic fields is the cyclotron energy $\hbar \omega_c$. The energy levels  depend through the SO index $\lambda$ on two other energy scales appearing in Eq. (\ref{LLpure}) via the dimensionless quantities $S$ and $Z$ which characterize the Rashba SO coupling and the Zeeman interaction, respectively. The explicit microscopic expressions for these quantities are provided in Sec. II.

An obvious effect of the SO coupling is to generate non-equidistant energy levels \eqref{LLpure}.
Moreover, the competition between the Zeeman and SO couplings leads to interesting effects for the energy spin splitting. 
 Indeed, two nearby energy levels with opposite index $\lambda$ can even become arbitrarily close and, at special values of the quantity
$S$ which depends on the magnetic field, the spin gap may vanish, giving rise to an accidental double degeneracy. These particular
degeneracy points have been previously related both to resonances in the spin
Hall conductance in the absence of disorder\cite{Shen2004,Shen2005,Shen2008} and to
the beating pattern of the DoS\cite{Morgenstern2010, Morgenstern2012}. 

It is worth emphasizing that the energy levels (\ref{LLpure}) in the pure case present also a large degeneracy with respect to the guiding center location (in other terms, the center of the cyclotron orbit), independently of the strengths of the Zeeman and SO couplings. 
All the degeneracies within the energy spectrum are expected to be lifted in the presence of a random potential energy.
In this work, we shall essentially address the associated fine structure of the  energy levels, i.e., we shall study in a quantitative way how the energy spectrum (\ref{LLpure}) is modified by the presence of an arbitrary  potential energy varying smoothly in space.
Note that, contrary to the DoS, a proper description of the LDoS behavior also requires to know in precise terms 
 the wave functions in addition to the energy spectrum. The theory developed in this paper shows how it is possible to devise a controlled approximate solution for the electronic states in the peculiar high magnetic field regime.

As a warming up, it is always instructive to consider toy models assuming a potential energy with a simple spatial dependence. Unfortunately, most of the simplest models are not tractable quantum mechanically in a fully analytical form and one has often to resort to numerical simulations to get some physical insight. For the hard-wall potential, there are theoretical studies using either a wave function formalism\cite{Grigoryan2009,Grigoryan2010}  or semiclassical approaches based on SU($2$) (spin) coherent states\cite{Reynoso2004}, with numerical studies mostly available in the literature\cite{Reynoso2004,Zhang2009,Pala2005}. The one-dimensional (1D) parabolic model for confinement is also not fully analytically tractable in the presence of an external magnetic field and both Rashba and Zeeman interactions. This  toy model for the edge states in the regime of the integer quantum Hall effect has been studied using numerical techniques\cite{Kramer2005} or analytically but without properly controlled approximation schemes\cite{Kushwaha2007}. Two-dimensional quadratic confining potentials have also been investigated, only numerically, as models for semiconductor quantum dots\cite{DasSarma2003, Emary2005, Chakraborty2006, Kushwaha2008,
Chakraborty2012}.

\section{Short Summary of the Results}
\subsection{Characteristic features of the quantum Hall regime}

The above-mentioned theoretical works have not specifically addressed the regime of the quantum Hall effect, for which the smooth disorder in the 2DEG plays a crucial role by producing both localized and delocalized electronic states. These disorder effects can be well captured within a semi classical picture\cite{Iordansky,Kaza,Trugman1983,Prange,Fogler1997}  involving a natural decomposition of the electronic motion in terms of a rapid cyclotronic motion and a slow drift of the guiding center. At high magnetic fields, these two kinds of motions decouple, constraining the guiding center to follow the equipotential lines of the disordered electrostatic potential. As a result, most of the states are localized, since the disordered potential landscape is principally constituted by closed equipotential lines. Delocalization of  the electronic states throughout the system is then only possible by following an extended percolating backbone occurring at a single critical energy and passing through many saddle points of the disorder landscape.\cite{Hashi}

A hallmark of this high magnetic field regime is thus the strong reduction of the communication between the cyclotronic and guiding center degrees of freedom, which corresponds to neglecting Landau level mixing in a quantum-mechanical picture. 
At the technical level, the projection within a single Landau level has been mainly presented in the literature\cite{Girvin} for the lowest Landau level, by exploiting the analyticity properties of the wave functions. The generalization of this wave function technique for the Landau levels $n \geq 1$ has been formulated \cite{Macris} 20 years later, at the price of numerous complications. In the recent years,  an alternative projection technique has been developed by two of us in the language of semi-coherent vortex Green's functions\cite{ChampelFlorens2008,ChampelFlorens2009,ChampelFlorens2010}. This vortex approach appears more general because it treats all the Landau levels on an equal footing, and allows one also, in principle, to include perturbatively the mixing between the Landau levels. It relies on a very basic idea, namely, the introduction of the orbital and guiding center degrees of freedom in the quantum realm by working preferentially with a basis of eigenstates  (in the pure case) characterized by two quantum numbers $n$ and ${\bf R}$. The degeneracy quantum number ${\bf R}=(X,Y)$ corresponds to the guiding center position in a classical picture and labels the location in the plane of the zeros of the wave function (via a coherent states algebra).  
The vortex theory is then nothing else but the translation in the quantum-mechanical language of the decomposition of the electronic motion into a fast cyclotronic rotation and a slow guiding center drift.

\begin{figure}[t]
\centering
\includegraphics[width=0.43\textwidth]{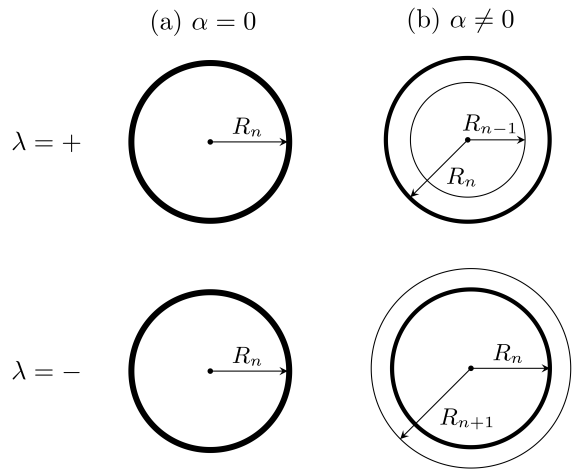}    
\caption{Schematic illustration of the quantum cyclotron orbits for the spin 
vortex states, column (a), and the SO vortex states for small SO coupling, column (b).
The top/bottom rows correspond to the different spin or SO indices
(for $\alpha=0$, one recovers the usual spin indices $\sigma$).
The probability density is peaked along two circles with cyclotron radii
$R_{n}$ and $R_{n\mp1}$ (the latter associated to the components $\sigma=+$ and $-$, 
respectively). 
The thick circles represent the dominant contribution to the probability
density, while the thin circles correspond to the sub-dominant one.
For non-zero SO coupling, each of the 
components of the spinor is sensitive to different averages of the disorder potential along 
the orbits, due to the differences in the probability density. This leads to a 
simple mechanism responsible for disorder potential-driven spatial fluctuations of
the energy spin splitting.} 
\label{fig:cyclotronorbits} 
\end{figure}

A substantial part of this paper is a generalization of this vortex approach by taking into account in the quantum-mechanical formalism both the charge and spin degrees of freedom, and, especially, their mutual coupling via the SO interaction. Indeed, a direct application of the previous results\cite{ChampelFlorens2008,ChampelFlorens2009,ChampelFlorens2010} is not possible {\em stricto sensu}, because the SO coupling leads to a specific spinorial form for the wave functions in the pure case, which requires the introduction of a new kind of vortex states.
From the geometric point of view, these SO vortex states labeled by the quantum numbers 
$(n,\mathbf{R},\lambda)$ present a probability density characterized by two
maxima peaked along the cyclotron orbits of two adjacent spin-resolved Landau levels
as represented in Fig. 1(b). The different radii collapse into a single one
$R_n$ in the limit of vanishing SO coupling as shown in Fig. 1(a) and
the spinor structure of the states becomes trivial. This already suggests a simple
physical picture to understand the role of disorder in quantum Hall systems with SO coupling: Each of the components of the spinor is sensitive to
different effective disorder potentials which result from the averaging of the potential energy along distinct cyclotron orbits,  and
the interplay between the two components gives the characteristic hallmarks of
quantum Hall systems with Rashba and Zeeman couplings. 

The theory developed here provides an  analytical derivation of this mechanism for disorder potential-driven spatial fluctuations of the  energy spin splitting in the framework of a semi-coherent states Green's function formalism.
As a simple estimation, we obtain in a weak SO coupling limit $S \ll |1-Z|$ in the case $Z <0$ (situation for InSb) an energy spin splitting [between the lowest-energy states $(1,+)$ and $(0,-)$] given by
 \begin{eqnarray}
\label{peaksgaps}
 E_s({\bf R}) \simeq E_{0,-}-E_{1,+} -\dfrac{1}{8}\left(\dfrac{S}{1-Z}\right)^2
l_B^2 \Delta_\mathbf{R}V(\mathbf{R}), \hspace*{0.5cm}
\end{eqnarray}
with $l_B$ the magnetic length and $\Delta_\mathbf{R}V(\mathbf{R})$ the
Laplacian of the potential energy function taken with respect to the guiding
center position. Expression (\ref{peaksgaps}) shows that nontrivial features appear due to
the interplay between Rashba SO coupling, Zeeman interaction and smooth
disorder. The resulting overall energy spin splitting appears well correlated with the disorder potential landscape, with a larger splitting obtained near the potential hills [where typically $\Delta_\mathbf{R}V(\mathbf{R}) <0$] than near the potential valleys [where $\Delta_\mathbf{R}V(\mathbf{R}) > 0$].

Smooth random fluctuations in the SO coupling parameter give rise to another mechanism for a spatial dispersion of the spin splitting, which will be also accounted for  within our Green's function formalism. The analytical expressions for the energy spectrum and the LDoS derived in this paper should be very helpful for a future thorough comparison between theory and STS experiments, in particular in order to quantify local fluctuations of Rashba SO coupling. We can nevertheless point out that the experimentally noticed\cite{Morgenstern2012} correlation between the spatial dispersion of the energy spin splitting and the local extrema of the potential landscape in InSb rather suggests that the contribution from the Rashba SO coupling fluctuations is seemingly less important than that induced by the spatial fluctuations of the disordered potential.

\subsection{Organization of the paper}

The paper is organized as follows. First, in Sec. \ref{spin_vortex_sec} we introduce the SO vortex states, which are peculiar eigenstates of the free electron Hamiltonian under perpendicular homogeneous magnetic fields in the presence of uniform Rashba and Zeeman interactions.  These states forming an overcomplete basis of semicoherent spinors constitute the elementary units for the developed theory. 
In Sec. \ref{Green_function_sec} we introduce the Green's function formalism using the SO vortex states and obtain the general equations of motion for the Green's function including energy level mixing processes in the
presence of disorder and fluctuations of the Rashba SO parameter. These equations, which can be related to a deformation quantization formulation of
quantum mechanics, are then solved in the limit of negligible coupling
between energy levels in Sec. \ref{Spectrum_sec} for electronic drift states. As a result, we obtain the energy spectrum for arbitrary locally flat potentials in the
presence of smooth fluctuations of the Rashba SO coupling parameter, the
formula becoming exact for globally flat 
potentials with
zero Gaussian curvature. 
Furthermore, we present in this Sec. \ref{Spectrum_sec}  simple analytical estimations of the spatial dispersion between
arbitrary spin-split energy sublevels. 
Finally, in Sec. \ref{LDoS_sec}, we use the Green's function obtained in Sec. \ref{Spectrum_sec} to analytically compute the LDoS in the quantum Hall regime in the presence of
a smooth arbitrary disorder, which can be described by locally flat potentials and
smooth Rashba fluctuations. This allows us to determine the spatial dispersion
and the linewidth of the LDoS peaks in different temperature regimes.

\section{Spin-Orbit Coupling in the Free Two-Dimensional Electron Gas}\label{spin_vortex_sec}
\subsection{Vortex states for the standard 2DEG}\label{vortex_subsec}

For the sake of simplicity, we first introduce the vortex states in the spinless case\cite{Champel2007}. They will be useful  to construct the elementary
components in the presence of SO coupling, upon which the whole Green's function theory relies.
We thus consider the Hamiltonian for a single spinless electron of effective mass $m^\ast$ and
electric charge $e=-|e|$ confined in a two-dimensional (2D) plane in the presence of
a  perpendicular magnetic field $\mathbf{B}=B\hat{\mathbf{z}}$: 
\begin{equation}\label{Hamiltonian2DEG}
  \hat{\mathcal{H}}_{\textnormal{2DEG}}=\dfrac{\hat{\mathbf{\Pi}}^2}{2m^{\ast}}=\dfrac{\hat{\Pi}_{x}^{2}+\hat{\Pi}_{y}^{2}}{2m^{\ast}},
\end{equation}
where 
\begin{equation}
  \hat{\mathbf{\Pi}}=-i\hbar \nabla_{\mathbf{r}}-\dfrac{e}{c}\mathbf{A}(\mathbf{r}),
\end{equation}
is the gauge-invariant momentum operator written in the position representation
[here $\mathbf{r}=(x,y)$ describes the position of the electron in the
2D plane and $c$ is the speed of light] and
$\mathbf{A}(\mathbf{r})$ is the electromagnetic vector potential related to the
magnetic field by the usual constitutive relation $\mathbf{B}=
\nabla_{\mathbf{r}} \times \mathbf{A}(\mathbf{r})$. The eigenvalue problem for
Hamiltonian \eqref{Hamiltonian2DEG}, $\hat{\mathcal{H}}_{\textnormal{2DEG}}
\Psi=E \Psi$, gives the well-known Landau spectrum characterized by discrete
energy levels
\begin{equation}
  E_{n}=\hbar \omega_{c}\left(n+\dfrac{1}{2}\right),
\end{equation}
with $n \geq 0$ a positive integer and $\omega_{c}=|e|B/(m^{\ast}c)$ the
cyclotron pulsation. The energy levels, labeled by the Landau level index
$n$, have a purely topological origin related to the quantization of the
magnetic flux enclosed by the cyclotron orbits induced by the Lorentz force on
the charged particles (or, correspondingly, the quantization due to
self-interference in the electronic circular motion). Accordingly, $n$ can be
interpreted as the number of magnetic flux quanta $\Phi_0=hc/|e|$ enclosed
by the cyclotron trajectory.

The Landau levels, $E_n$, are infinitely degenerate since the motion of the
electron has two degrees of freedom (we thus expect here two quantum numbers to be
involved). This means that
there is a great liberty in the choice of the basis states which diagonalize the
Hamiltonian in Eq. \eqref{Hamiltonian2DEG}, the particular choice depending on
the symmetry of the gauge-invariant probability density $|\Psi|^2$. Imposing the
probability density to be a function of $|\mathbf{r}-\mathbf{R}|$ only, so that
it reflects the classical orbital motion of the electron around a guiding center ${\bf R}$, we obtain the set of overcomplete vortex
states\cite{Malkin, Champel2007} which are expressed in the
symmetrical gauge $\mathbf{A}(\mathbf{r})=\mathbf{B} \times \mathbf{r}/2$ as
\begin{multline}\label{vortex}
  \Psi_{n,\mathbf{R}}(\mathbf{r})=\dfrac{1}{\sqrt{2 \pi l_{B}^{2} n!}} 
\left[\dfrac{x-X+i(y-Y)}{\sqrt{2}l_{B}} \right]^{n} \\ \times  
\exp \left[-\dfrac{(x-X)^2+(y-Y)^2+2i(yX-xY)}{4l_{B}^{2}}\right],
\end{multline}
where $l_B=\sqrt{\hbar c/(|e|B)}$ is the magnetic length.

The vortex states, which can be written as
$\Psi_{n,\mathbf{R}}(\mathbf{r})=\langle \mathbf{r} |n,\mathbf{R} \rangle$  in the Dirac notation, are
characterized by the set of quantum numbers $\nu=\{n,\mathbf{R}\}$. They are so called because the continuous
quantum number $\mathbf{R}=(X,Y)$ characterizes (for $n\geq1$) the position of the zeros of the wave function, which  corresponds to
vortex-like phase singularities in the 2D plane. The eigenstates $|n,{\bf R} \rangle$ form a peculiar semicoherent basis, since they satisfy  the coherent states algebra with respect to the
continuous (degeneracy) quantum number $\mathbf{R}$.  As a consequence, they are orthogonal with respect to the Landau level
index but non-orthogonal with respect to the vortex position. They form a
semiorthogonal basis with the overlap
\begin{equation}\label{semiorthogonality_vortex}
  \langle n_{1},\mathbf{R}_{1}|n_{2},\mathbf{R}_{2}\rangle=\delta_{n_1,n_2}\langle \mathbf{R}_{1}|\mathbf{R}_{2} \rangle, 
\end{equation}
where
\begin{equation}\label{overlap_vortex}
  \langle \mathbf{R}_{1}|\mathbf{R}_{2}\rangle=\exp{\left[ - \dfrac{(\mathbf{R}_{1}-\mathbf{R}_{2})^2-2i \hat{\mathbf{z}} \cdot(\mathbf{R}_{1}\times\mathbf{R}_{2})}{4 l^2_{B}}\right]}.
\end{equation}
Importantly,  they obey the following completeness relation
\begin{equation}\label{completeness_vortex}
 \int \dfrac{d^{2} \mathbf{R}}{2 \pi l^2_{B}} \sum_{n=0}^{+\infty}|n,\mathbf{R}\rangle\langle n,{\mathbf{R}}|=1\!\!1_{\textnormal{orb}},
\end{equation}
which  allows us to project the 2D electron dynamics within this vortex
representation. 

The vortex basis also provides considerable advantages in order to describe the lifting
of the energy degeneracy by an arbitrary (smooth) potential landscape $V(\mathbf{r})$,
since the degeneracy quantum number does not result from  a particular symmetry, in contrast to the Landau states basis expressing a translation invariance, or the circular states basis characterized by a global rotation invariance.
More precisely, the degeneracy in the vortex representation is grasped from a differential geometry perspective via the continuous position ${\bf R}$, which avoids to define the specific shape of the quantum cell. As a result, the vortex states are characterized by a great local adaptability to random spatial
variations of the potential energy, i.e., they display some robustness properties in response to arbitrary local perturbations. This is the basic reason why they are chosen as a preferred set of states to deal with a realistic description of disorder effects.

\subsection{Spin-orbit vortex states}\label{spinvortex}
Now, we consider that the electron has a spin
$s=\frac{1}{2}$. The single particle Hamiltonian
for the electron in the presence of Rashba SO coupling and Zeeman
interaction can then be written as
\begin{equation}\label{Hamiltonian_0}
  \hat{\mathcal{H}}_0=\hat{\mathcal{H}}_{\textnormal{2DEG}}\otimes 1\!\!1_\textnormal{s} 
+ \hat{\mathcal{H}}_{\textnormal{R}}+ \hat{\mathcal{H}}_{\textnormal{Z}},
\end{equation}
where $\hat{\mathcal{H}}_{\textnormal{2DEG}}$ is given in Eq.
\eqref{Hamiltonian2DEG} (here $1\!\!1_\textnormal{s}$ is the $2 \times 2 $
identity matrix that represents the identity operator in spin space and
$\otimes$ the tensor product symbol). $\hat{\mathcal{H}}_{\textnormal{R}}$ is
the Rashba Hamiltonian\cite{Rashba1960} which describes the coupling between the
orbital and spin degrees of freedom
\begin{equation}\label{RashbaHamiltonian}
  \hat{\mathcal{H}}_\textnormal{R}=\alpha [\hat{\mathbf{\Pi}} \times 
\bm{\sigma}]_z=\alpha [\hat{\Pi}_x \otimes \sigma_y - \hat{\Pi}_y \otimes \sigma_x],
\end{equation}
with $\alpha \equiv \langle \alpha({\bf r}) \rangle$ the spatially averaged Rashba
SO parameter and $\bm{\sigma}=(\sigma_x,\sigma_y,\sigma_z)$ a vector
whose components are the Pauli matrices. Finally,
$\hat{\mathcal{H}}_{\textnormal{Z}}$ is the Zeeman interaction term
\begin{equation}
  \hat{\mathcal{H}}_{\textnormal{Z}}=\dfrac{1}{2}g \mu_{B} B \otimes\sigma_{z},
\end{equation}
which describes the coupling between the electron's spin and the external
magnetic field. Here, $g$ is the Land\'e $g$ factor and $\mu_B=|e|\hbar/(2 m_0
c)$ is the Bohr's magneton with $m_0$ the bare electron mass. 

Since the Hamiltonian in Eq. \eqref{Hamiltonian_0} presents a matrix structure
in the spin space, 
we shall look for wave functions solutions of the eigenvalue problem 
\begin{equation}\label{eigenvalue_0}
  \hat{\mathcal{H}}_0 \tilde{\Psi}=E \tilde{\Psi},
\end{equation} with the following SO vortex states
\begin{equation}\label{spinvortex_states}
 \tilde{\Psi}_{n,\mathbf{R}}(\mathbf{r})
=\sum_{\sigma=\pm}f_{\sigma}(\theta)\Psi_{n_\sigma,\mathbf{R}}(\mathbf{r}) \otimes |\sigma\rangle,
\end{equation}
where $\Psi_{n_\sigma,\mathbf{R}}(\mathbf{r})$ are the spinless vortex states given in
Eq. \eqref{vortex}, $|\sigma \rangle$ are the eigenstates of the Pauli matrix
$\sigma_z$, i.e., $\sigma_z |\sigma \rangle=\sigma |\sigma \rangle$, and
 the weights $f_\sigma(\theta)$ of the spinor components are defined according to
\begin{equation}\label{weight_f}
 f_\sigma(\theta)=
\begin{cases}
 \sin (\theta) & \sigma=+,\\
\cos (\theta) & \sigma=-,
\end{cases}
\end{equation}
and
\begin{equation}\label{nsigma}
 n_\sigma=
\begin{cases}
n-1 & \sigma=+,\\
n & \sigma=-.         
\end{cases}
\end{equation}
This form of the spinor wave function can be traced back to Eq. \eqref{RashbaHamiltonian} written in terms of the matrices $\sigma_\pm=\sigma_x \pm i\sigma_y$, where it is easy to see that the Rashba Hamiltonian couples Landau levels which differ in just one unit. The diagonalization is straightforward by defining the  operators $\hat{\Pi}_\pm\equiv\hat{\Pi}_x \pm i\hat{\Pi}_y$ whose action on the vortex states is 
\begin{subequations}
  \begin{align}
 \hat{\Pi}_{+}|n,\mathbf{R} \rangle&=i\hbar \dfrac{\sqrt{2}}{l_B}\sqrt{n+1}|n+1,\mathbf{R} \rangle,\\
 \hat{\Pi}_{-}|n,\mathbf{R} \rangle&=-i\hbar \dfrac{\sqrt{2}}{l_B}\sqrt{n}|n-1,\mathbf{R} \rangle.
  \end{align}
\end{subequations}
Performing the substitution into Eq. \eqref{eigenvalue_0}, we get the following set of coupled algebraic linear equations
\begin{subequations}
\begin{align}
\left(E_{n-1}+\dfrac{1}{2}g \mu_B B \right)-\alpha \hbar \dfrac{\sqrt{2n}}{l_B} \cot\theta &=E, \\
\left(E_{n}-\dfrac{1}{2}g \mu_B B \right) - \alpha \hbar \dfrac{\sqrt{2n}}{l_B} \tan\theta &=E.
\end{align}
\end{subequations}
which can be readily solved. The eigenenergies for the Hamiltonian of the clean system are therefore
\begin{equation}\label{eigenenergies_0}
  E \equiv E_{n,\lambda}=\hbar \omega_{c}\left[n-\dfrac{\lambda}{2}\sqrt{(1-Z)^{2}+nS^{2}} \right],
\end{equation}
with $n \geq 1$ and $\lambda = \pm$ the SO quantum number \cite{footnote1}. For $n=0$ the above equation still holds but we have necessarily $\lambda=-$. The dimensionless parameters $S$ and $Z$, which measure the strength of the Rashba SO coupling (per magnetic length) and the Zeeman interaction relative to the cyclotron energy,  are defined as
\begin{align}
 S&\equiv\dfrac{\alpha 2 \sqrt{2}}{\omega_c l_B}, \label{S} \hspace*{1cm}
 Z\equiv\dfrac{g \mu_B}{\hbar |e|}m^\ast c= \dfrac{g}{2} \dfrac{ m^{\ast}}{ m_0}.
\end{align}
We shall assume throughout this paper that $1-Z >0$.

 As in the spinless 2DEG described in Sec.
\ref{vortex_subsec}, the eigenenergies are again highly degenerate with respect to the guiding center position in the absence of potential energy. This means that
there is a great freedom in the choice of the basis which diagonalizes Eq.
\eqref{Hamiltonian_0}, this liberty being already taken into account in the particular
(non-unique) choice of the ansatz \eqref{spinvortex_states}.

The energy spectrum
\eqref{eigenenergies_0}  formally interpolates between the
usual Landau spectrum, linear in the Landau level index and given by $E_{n,\pm}
= \hbar \omega_c (n+1/2\pm Z/2)$ for $S=0$, and the relativistic (graphene-like)
spectrum, with its characteristic square-root dependence on $n$ expected for massless Dirac fermions, $E_{n,\pm} \simeq \pm \hbar \Omega_c
\sqrt{n}$ for $n\geq 1$ in the limit $|S|\gg\sqrt{n}$ (here, $\Omega_c = \alpha
\sqrt{2}/l_B$ can be identified with the
graphene characteristic frequency once we recognize that $\alpha$ plays the same
role as the Fermi velocity $v_\textnormal{F}$).  The limit $|S| \rightarrow 0$, for which the SO and the spin quantum numbers become equivalent
$\lambda\equiv\sigma$, requires a relabeling of the energy level index according to
the mapping $n_{-\sigma} +1 \rightarrow n$, in order to reintroduce the picture of the splitting of each Landau level into two spin-polarized sublevels by the Zeeman interaction only.

In addition, energy spectrum \eqref{eigenenergies_0} is quite rich and presents multiple level crossings as a function of the SO coupling $\alpha$ or the magnetic field. These crossings occur whenever
$E_{n_1,\lambda_1}=E_{n_2,\lambda_2}$ with $n_1 \neq n_2$ and necessarily $\lambda_1 =- \lambda_2$, a condition which can be
cast in the form of a biquadratic equation for the dimensionless parameter $S$
\begin{equation}
  S^{4}-8(n_1+n_2)S^{2}+16[(n_1-n_2)^{2}-(1-Z)^{2}]=0.
\end{equation}
This equation yields level intersections for the special values 
\begin{equation}\label{resonance_condition}
  S_c= 2\sqrt{(n_1+n_2)- \sqrt{4n_1 n_2+(1-Z)^{2}}}.
\end{equation}
We deduce that the crossings involve different energy levels such that
$|n_1-n_2|>1-Z$. The associated double degeneracy is expected to be lifted in the presence of a smooth disorder potential.

\begin{figure}
\centering
\includegraphics[width=0.43\textwidth]{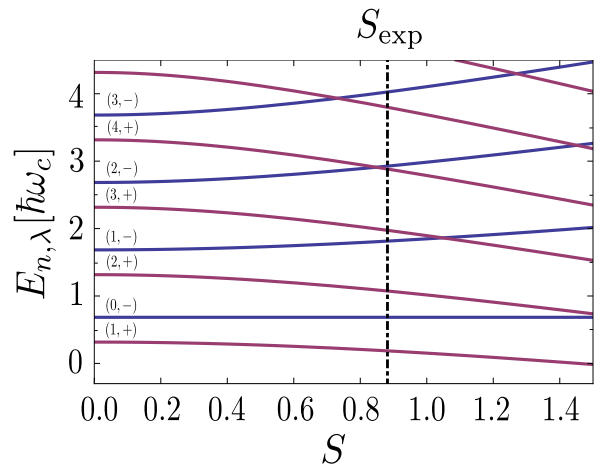}
\caption{ (Color online)
Energy spectrum (in units of the cyclotron energy $\hbar \omega_c$) resulting from Eq. \eqref{eigenenergies_0} in the absence 
of potential energy, as a function of the dimensionless SO 
parameter $S$. Values are taken  from STS measurements~\cite{Morgenstern2010,Morgenstern2012} in InSb semiconductor: 
$m^\ast = 0.035 m_0$, $g = -21$, $\hbar \alpha = 7 \cdot 10^{-11}$ eVm.
The dashed line shows the particular value $S_{\mathrm{exp}} \simeq 0.88$ (reached for $B=7$ T) at which a pronounced spatial dispersion of the energy spin splitting  has been noticed experimentally\cite{Morgenstern2012}.
}
\label{fig2} 
\end{figure}
Physically speaking, it is always interesting to have an idea of the energy scales involved. As a general rule, the characteristic Rashba energy
$E_{\mathrm{so}} = m^\ast \alpha^2$ is of the order of $0.1-1.0$ meV, one order of magnitude
below the typical energy scale related to cyclotron motion (at $B=1$ T). However, in 2D
heterostructures where the SO coupling is strong due to heavy elements
such as in InSb, both energies can be of similar order of magnitude. For
example, considering the values of the effective mass $m^\ast=0.035m_0$, the
Land\'e $g$ factor $g \simeq -21$ and the Rashba coupling constant $\hbar \alpha = 7 \cdot 10^{-11}$ eVm taken from Ref. \onlinecite{Morgenstern2010}, the
cyclotron energy at $B=1$ T is $\hbar \omega_c \simeq 3$ meV,  which is of the
same order as the Rashba characteristic energy ($E_{\mathrm{so}} \simeq
3$ meV). In the clean spectrum \eqref{eigenenergies_0} the
relevant quantities  are the Rashba and
Zeeman dimensionless parameters $S$ and $Z$ given by Eq. \eqref{S},
which take the values $S \simeq 2.33$ and $Z\simeq -0.37$ for $B=1$ T. We have plotted in Fig. \ref{fig2} the resulting energy spectrum (\ref{eigenenergies_0}).
Note that, in the absence of electron-electron interaction, the quantity $S$ decreases when increasing the magnetic field amplitude
 as $S \sim l_B$, while $Z$ on the other hand remains
constant. Since even for high magnetic fields of several Teslas,
$S$ can be of the order of unity, the understanding of the interplay between
Zeeman and Rashba couplings in a given energy level $n$ becomes crucial.

Using the Dirac notation and introducing the multi-index $\nu
=\{n,\mathbf{R},\lambda \}$ for the set of quantum numbers, the normalized SO vortex states take the form
\begin{equation}\label{SO_vortex_states}
|\nu \rangle\equiv|n,\mathbf{R},\lambda \rangle =  
\sum_{\sigma=\pm}f_\sigma(\theta_n^\lambda)|n_\sigma,\mathbf{R}\rangle \otimes |\sigma \rangle,
\end{equation}
where the angles $\theta_n^\lambda$ are defined by 
\begin{equation}\label{angles}
 \theta^{\lambda}_{n}= \arctan \left[ \dfrac{(1-Z)+\lambda\sqrt{(1-Z)^2 + n S^2}}{S \sqrt n} \right],
\end{equation}
for $n \geq 1$ and $\theta^{-}_{0}=0$ if $n=0$. For $n\geq 1$ we also have the relation
\begin{equation}\label{relationorthogonality}
 \theta^{+}_{n}=\theta^{-}_{n}+\pi/2,
\end{equation}
which guarantees the orthogonality of the SO vortex states belonging to
the same level $n$ but having opposite SO quantum number $\lambda$.
Furthermore, Eq. \eqref{relationorthogonality} implies that the function
$f_\sigma(\theta_n^\lambda)$ satisfies the following sum rules
\begin{subequations}
\begin{align}\label{relation_f_1}
\sum_{\sigma=\pm}f_\sigma(\theta_n^{\lambda_1})f_\sigma(\theta_n^{\lambda_2})&=\delta_{\lambda_1,\lambda_2},\\
\sum_{\lambda=\pm}f_{\sigma_1}(\theta_n^{\lambda})f_{\sigma_2}(\theta_n^{\lambda})&=\delta_{\sigma_1,\sigma_2}\label{relation_f_2},
\end{align}
\end{subequations}
which can be seen as completeness relations that hold in the $\lambda$ and $\sigma$ subspaces.

Not surprisingly, the SO vortex states present the same properties as their spinless counterparts. Using  the orthonormality
relation satisfied by the spin states $\langle \sigma_1|\sigma_2
\rangle=\delta_{\sigma_1,\sigma_2}$, it can be readily checked that the
SO vortex states are semiorthogonal
\begin{align}\label{semiorthogonality_spinvortex}
 \langle \nu_{1} | \nu_{2}\rangle&=\sum_{\sigma_1,\sigma_2} f_{\sigma_1}(\theta_{n_1}^{\lambda_1}) f_{\sigma_2}(\theta_{n_2}^{\lambda_2}) \notag 
\langle n_{1\sigma_1},\mathbf{R}_1|n_{2\sigma_2} ,\mathbf{R}_2\rangle \langle \sigma_1|\sigma_2 \rangle,\\ \notag
 &= \delta_{n_1,n_2} \langle \mathbf{R}_{1}|\mathbf{R}_{2} \rangle \sum_{\sigma_1=\pm} f_{\sigma_1}(\theta_{n_1}^{\lambda_1}) f_{\sigma_1}(\theta_{n_2}^{\lambda_2}),  \\ &= \delta_{n_1,n_2} \langle \mathbf{R}_{1}|\mathbf{R}_{2} \rangle \delta_{\lambda_1,\lambda_2}.
\end{align}
Finally, introducing the short-hand notation
\begin{equation}\label{sum_quantumnumbers}
  \sum_{\nu}= \int \dfrac{d^2 \mathbf{R}}{2 \pi l_B^2}\sum_{n=0}^{+\infty} \sum_{\lambda=\pm},
\end{equation}
for the sum over the quantum numbers and using the completeness relation
satisfied by the vortex states \eqref{completeness_vortex}, we can easily verify that the set of
SO vortex states forms a basis with the completeness relation
\begin{align}\label{completeness_spinvortex}
  \sum_{\nu} |\nu \rangle \langle \nu| &= \notag \int \dfrac{d^{2} \mathbf{R}}{2 \pi l^2_{B}} \sum_{n=0}^{+\infty} \sum_{\lambda=\pm} \sum_{\sigma=\pm} \sum_{\sigma'=\pm}f_\sigma(\theta_n^\lambda)f_{\sigma'}(\theta_n^\lambda) \\ \notag&\times |n_\sigma,\mathbf{R} \rangle \langle n_{\sigma'},{\mathbf{R}}| \otimes |\sigma \rangle \langle \sigma'|, \\ &=\notag \int \dfrac{d^{2} \mathbf{R}}{2 \pi l^2_{B}} \sum_{n=0}^{+\infty}  |n,\mathbf{R} \rangle \langle n,{\mathbf{R}}|  \otimes \sum_{\sigma=\pm} |\sigma \rangle \langle \sigma |,\\& =1\!\!1_\textnormal{orb} \otimes  1\!\!1_\textnormal{s} \equiv 1\!\!1.
\end{align}

\section{Green's Function Formalism for Disordered Quantum Hall Systems with Rashba Spin-Orbit Coupling}\label{Green_function_sec}

\subsection{Disorder and fluctuations of spin-orbit coupling} 
We consider now that the electron feels, in addition to the external perpendicular magnetic field, the presence of a (generalized) potential. Therefore, the Hamiltonian will contain, besides the kinetic energy part $\hat{\mathcal{H}}_0$ given in Eq. \eqref{Hamiltonian_0}, a potential energy term $\hat{U}$
\begin{equation}\label{Hamiltonian}
 \hat{\mathcal{H}}=\hat{\mathcal{H}}_0+\hat{U}.
\end{equation}
The operator $\hat{U}$ can be written as
\begin{equation}\label{generalized_potential}
 \hat{U}=\hat{V}(\mathbf{r})\otimes1\!\!1_\textnormal{s} + \delta \hat{\mathcal{H}}_{\textnormal{R}},
\end{equation}
with $V(\mathbf{r})$ a scalar potential and $\delta \hat{\mathcal{H}}_{\textnormal{R}}$ the fluctuating Rashba Hamiltonian operator
\begin{equation} 
\delta \hat{\mathcal{H}}_{\textnormal{R}}= \dfrac{1}{2}\left\{ \widehat{\delta\alpha}(\mathbf{r}),[\hat{\mathbf{\Pi}} \times \bm{\sigma}]_z\right\}.
\end{equation}

Here $\{\,\cdot \,,\,\cdot\,\}$ is the anticommutator [i.e., $\{\hat{A},\hat{B}\}\equiv \hat{A}\hat{B}+\hat{B}\hat{A}$ where $\hat{A}$ and $\hat{B}$ are two arbitrary operators] which ensures the Hermiticity of the fluctuating Rashba Hamiltonian and accounts for the noncommutativity between the spatial fluctuations  of the Rashba parameter $\delta \alpha(\mathbf{r})$ and the gauge-invariant momentum. These spatial fluctuations are induced by random local electric fields perpendicular to the 2DEG plane, fluctuations in the concentration of donor ions or randomness in the direction of the crystal axis due to inhomogeneous growth or local strain\cite{Sherman2003, Sherman2005}.
In principle, we shall require the correlations $\langle \delta \alpha(\mathbf{r}) \delta \alpha(\mathbf{r}') \rangle$ of the Rashba SO coupling parameter to be described by a smooth distribution function which depends only on the difference between two electronic positions $\mathbf{r}-\mathbf{r}'$ (this correlation function is {\em a priori} different from the correlation function that characterizes the spatial fluctuations of the scalar potential).

The scalar potential $\hat{V}(\mathbf{r})$ in Eq. \eqref{generalized_potential} accounts for several physical mechanisms: it includes the effect of confinement, random impurity potentials, mean-field Coulomb interaction between the electrons or external non-equilibrium electric fields. This potential can be strikingly different from the bare electrostatic one due to screening effects, i.e. redistribution of the electron density at the Fermi level, leading to the formation in the sample of alternating compressible and incompressible regions of different widths at high magnetic fields\cite{Glazman1992, Chalker1993}. Note that, in principle, it is necessary to include both direct and exchange interactions between electrons in order to microscopically determine the total scalar potential\cite{Stoof1995,Larkin2000,Manolescu}.

In addition, the exchange coupling can renormalize (and enhance) the Rashba SO interaction parameter\cite{Raikh1999} and the Land\'e $g$ factor\cite{Piot} in 2DEG. This enhancement of Rashba SO interaction can be described within the present theory since it can be included as an additional fluctuation $\delta \alpha$ of the bare SO coupling parameter $\alpha$. In the case of the Land\'e $g$ factor, our theory also accounts for a global enhancement by simply replacing the bare $g$ factor by a renormalized one $g^\ast$ which now can depend on the external parameters such as the magnetic field, temperature or the macroscopic electron density. Local enhancement of the $g$ factor requires a minor modification to this  theory (not presented here) where the spin-diagonal scalar potential is substituted by a scalar potential that depends on the spin projection $\hat{V}_\sigma (\mathbf{r})$.   

\subsection{Equation of motion for the Green's function in the spin-orbit vortex representation} 
To  investigate the combined effects of a smooth disorder potential and random Rashba fluctuations on the electron dynamics, we shall use a semicoherent Green's function formalism which was previously developed to study disordered 2DEG (Ref. \onlinecite{ChampelFlorens2009}) and graphene\cite{ChampelFlorens2010} in the quantum Hall regime.  

The Green's operators associated to the Hamiltonian \eqref{Hamiltonian} 
are defined by the equation 
\begin{equation} 
 (\omega- \hat{\mathcal{H}}\pm i0^+)\hat{G}^{R,A}(\omega)=1\!\!1,\label{Green_operator}
\end{equation} 
where the plus (minus) sign corresponds to the retarded (advanced) Green's operator, $0^+$ is a positive infinitesimal quantity which encodes the information about the boundary conditions for time evolution and $1\!\!1$ is the identity operator.
The projection of the operator equation \eqref{Green_operator} onto a given basis of states yields the equation of motion for the Green's function  written in the energy representation (here $\omega$ indicates the energy resulting from the Fourier transformation of the relative time dependence $t_1-t_2$ of the Green's function).  
Alternatively, we may also introduce the Green's function in terms of the field operators $\hat{\psi}(x)$ [evaluated at a given point of the space-time $x=(\mathbf{r},t)$] in the electronic representation 
\begin{eqnarray} 
 G^{R,A}(x_1;x_2) = \mp i\Theta \left[\pm(t_{1}-t_{2}) \right] \langle\{\hat{\psi}(x_1),\hat{\psi}^{\dagger}(x_2)\}\rangle,  \hspace*{0.5cm}
\end{eqnarray} 
where $\Theta(t)$ is the 
Heaviside step function [$\Theta(t)=0$ if $t<0$ and $\Theta(t)=1$ if $t\geq 0$] 
and the brackets $\langle \cdot \rangle$ represent the thermodynamic average in 
the grand-canonical ensemble. 
Because we consider a time-independent Hamiltonian, the energy is conserved and the Green's functions only depend   on the time difference 
$\tau=t_1-t_2$. 

The completeness relation \eqref{completeness_spinvortex} satisfied by the SO vortex basis
allows us to express the Green's operator in the SO vortex representation 
$\{|\nu \rangle \}$. Within this representation, the Green's function 
$G_{\nu_1;\nu_2}^{R,A}(\omega)=G^{R,A}(n_1,\lambda_1,\mathbf{R}_1;n_2,\lambda_2,\mathbf{R}_2;\omega)$ 
gives the probability amplitude for a vortex with circulation $n_1$ and 
SO quantum number $\lambda_1$ located at a position $\mathbf{R}_1$ to be
scattered elastically (energy $\omega$ is conserved within the process) at a position $\mathbf{R}_2$ with the new circulation $n_2$ and SO
quantum number $\lambda_2$. In the absence of potential ($\hat{U}=0$), the SO vortex states are eigenstates of the Hamiltonian $\hat{\mathcal{H}}_0$, so that  from the projection of Eq. \eqref{Green_operator} we get straighforwardly  the  unperturbed  Green's function
\begin{equation} 
 G_{0\,\nu_1;\nu_2}^{R,A}(\omega)=\dfrac{\delta_{n_1,n_2}\langle \mathbf{R}_1| \mathbf{R}_2 \rangle \delta_{\lambda_1,\lambda_2}}{\omega-E_{n_1,\lambda_1}\pm i0^+}, 
\end{equation} 
where $G^{R,A}_{0\,\nu_1;\nu_2}(\omega)=\langle
\nu_1|\hat{G}_{0}^{R,A}(\omega)|\nu_2 \rangle$ is the kernel of the free Green's 
operator [here, we have used the semiorthogonality property 
\eqref{semiorthogonality_spinvortex} of the SO vortex states]. In the clean case, the Green's function  is therefore diagonal both in the 
level index $n$ and the SO quantum number, and  presents the typical coherent states nonzero overlap for the vortex position dependence. 
 
In the presence of a potential $\hat{U}$ (the particular form of 
the potential is not important at this point), the Green's function can, in principle, be 
obtained by solving the Dyson equation which can be written from projecting Eq. \eqref{Green_operator}  onto the SO vortex basis and using the completeness relation \eqref{completeness_spinvortex} 
\begin{multline}\label{Dyson_general_energy} 
  (\omega-E_{n_1,\lambda_1}\pm i0^{+})G^{R,A}_{\nu_1;\nu_2}(\omega)\\=\langle \nu_{1}|\nu_{2}\rangle + \sum_{\nu_{3}}U_{\nu_{1};\nu_{3}}G_{\nu_3;\nu_2}^{R,A}(\omega), 
\end{multline} 
where $U_{\nu_1;\nu_2}=\langle \nu_1|\hat{U}|\nu_2 \rangle$ are the matrix 
elements for the potential $\hat{U}$ in the SO vortex representation.  It is clear
from Eq. \eqref{Dyson_general_energy} that whenever $\hat{U} \neq 0$, the
Green's function will generally be no longer diagonal with respect to the discrete quantum 
numbers $n$ and $\lambda$, the mixing between the latter depending on the 
particular form of the potential energy function $\hat{U}$.  In the following, we shall only concentrate on the determination of the retarded Green's function, given that the advanced one can be trivially inferred from the knowledge of the retarded function at equilibrium. In order not to burden the expressions unnecessarily, we shall also drop the retarded superscript.

\subsection{Mixed phase space formulation of Dyson equation in the spin-orbit vortex representation} 
Solving analytically Dyson equation \eqref{Dyson_general_energy} for an arbitrary potential $\hat{U}$ is a very difficult task. Nevertheless, it has been found in Refs. 
\onlinecite{ChampelFlorens2008,ChampelFlorens2009} that, as a result of the 
coherent-state character of the degeneracy quantum number $\mathbf{R}$, the matrix 
elements of the potential and the Green's function must necessarily take the form 
\begin{align} 
  U_{\nu_1;\nu_2}&= \langle \mathbf{R}_{1}| \mathbf{R}_{2} \rangle T_{\mathbf{R}_{12}} \left[u_{n_1,\lambda_1;n_2,\lambda_2}(\mathbf{R}_{12})\right], \label{V_vortex}\\ 
  G_{\nu_1;\nu_2}(\omega)&=\langle \mathbf{R}_{1}| \mathbf{R}_{2}\rangle T_{\mathbf{R}_{12}} \left[g_{n_1,\lambda_1;n_2,\lambda_2}(\mathbf{R}_{12},\omega)\right],\label{Green_function_nonlocal} 
\end{align} 
where the vortex overlap $\langle \mathbf{R}_{1}| \mathbf{R}_{2} \rangle$  which contains the non-analytical dependence on the magnetic length has been extracted. Here, $T_{\mathbf{R}}$ represents the differential Gaussian operator defined as 
\begin{equation}\label{Gaussian_exp} 
 T_{\mathbf{R}}\equiv\exp\left(\dfrac{l^{2}_{B}}{4}\Delta_{\mathbf{R}}\right), 
\end{equation} 
with $\Delta_{\mathbf{R}}$ the Laplacian taken with respect to the vortex 
position and 
$\mathbf{R}_{12}=[\mathbf{R}_1+\mathbf{R}_2+i(\mathbf{R}_1-\mathbf{R}_2) \times
\hat{\mathbf{z}}]/2$ a particular (complex) combination of the center of mass and relative 
coordinates of two vortex positions. This functional dependence  implies that the full nonlocal Green's function is 
completely specified once the local SO vortex 
Green's function $g_{n_1,\lambda_1;n_2,\lambda_2}(\mathbf{R},\omega)$ at 
coinciding vortex positions $\mathbf{R}_1=\mathbf{R}_2 \equiv\mathbf{R}$ is known. In fact, this diagonal representation with respect to the vortex position is a well-known property of coherent states\cite{Sudarshan}.

After using the forms \eqref{V_vortex} and \eqref{Green_function_nonlocal} and following the same steps as for the standard 2DEG without 
SO coupling\cite{ChampelFlorens2008}, 
Eq. 
\eqref{Dyson_general_energy} can  be exactly mapped onto the following equation 
of motion for the function $g(\mathbf{R},\omega)$: 
\begin{multline}\label{Dyson_star_1} 
  (\omega-E_{n_1,\lambda_1} + i0^+) {g}_{n_1,\lambda_1;n_2,\lambda_2} (\mathbf{R},\omega) = \delta_{n_1,n_2}\delta_{\lambda_1,\lambda_2} \\ + \sum_{n_3,\lambda_3} {u}_{n_1,\lambda_1;n_3,\lambda_3}(\mathbf{R}) \star {g}_{n_3,\lambda_3;n_2,\lambda_2}(\mathbf{R},\omega). 
\end{multline}
The symbol $\star$ represents the pseudodifferential infinite order symplectic operator defined as 
\begin{equation}\label{starproduct} 
\star \equiv \exp{\left[i \dfrac{l^{2}_{B}}{2}(\overleftarrow{\partial}_{X}  \overrightarrow{\partial}_{Y}-\overleftarrow{\partial}_{Y} \overrightarrow{\partial}_{X}) \right]},
\end{equation} 
where the arrow above the partial derivative indicates to which of the factors (left/right) the partial derivative is applied.  
In this form, Eq. \eqref{Dyson_star_1} is still a complicated matrix partial differential equation of infinite order. However, as we shall see in Sec. \ref{Spectrum_sec}, it can be solved in high magnetic fields for important cases depending on the specific form of the potential energy function. 

The $\star$-product defined in Eq. \eqref{starproduct} is completely analogous 
to the Groenewold-Moyal product (see for instance Ref. \onlinecite{Zachos2002}), with $l^2_B$ playing the role of an effective Planck's constant and the 1D conjugated variables $(x,p_x)$ being replaced by $(X,Y)$.
It is ubiquitous in the 
deformation quantization theory\cite{Zachos2002}, an alternative formulation of quantum mechanics in phase space, where the central object is the Wigner function in place of the wave function. 
In this context, the $\star$-product allows to 
express quantum laws for non-commuting quantum operators in terms of commuting 
variables making the correspondence between classical and quantum substrates 
more transparent than in the Hilbert space approach, since classical mechanics is obtained smoothly by a continuous limit of the deformation parameter, $l_B 
\rightarrow 0$. This non-commutative product between functions 
can also be found in string theory, spin field theory and, in general, in 
noncommutative field theory\cite{Douglas2001}. Transposed to the 2DEG under 
perpendicular magnetic fields, this formulation becomes\cite{ChampelFlorens2010} a mixed phase space 
deformation quantization theory that combines discrete Landau levels and a 
continuous phase space, which corresponds to the real space for the guiding center coordinates $\mathbf{R}=(X,Y)$.  
 
In the framework of the deformation quantization theory, we can also give a meaning to the operator $T_{{\bf R}}$ defined in Eq. \eqref{Gaussian_exp}: It is simply the (invertible) transition operator which dresses  the so-called Wick-Voros product that controls the dynamics of Husimi functions\cite{Zachos2000}
into
the Groenewold-Moyal product.  The Husimi function can be directly defined as the trace of the density matrix over the basis of coherent states and turns out to be a Gaussian-smoothed Wigner function. 
Both products originate from a generalized Weyl map \cite{Zachos2000,Vitale2008} which associates phase space 
functions to operators according to certain quantization rules (Weyl or 
symmetric and Wick or normal order, respectively). Although the passage from the 
Wick-Voros to the Groenewold-Moyal product can be regarded\cite{Vitale2008} as a 
trivial rotation in phase space, it has highly non-trivial consequences in the present case, since it allows one to tackle rather easily the Dyson equation in the SO vortex representation in the case of 1D potentials $\hat{U}({\bf r})$ (edge states problem), see further. In some way, the operator $T_{{\bf R}}$ realizes the delocalization of the SO vortex states (which are originally localized in any direction) along the equipotential lines of $\hat{U}({\bf r})$. As a final remark, it is worth specifying that we are actually dealing here with peculiar Green's functions rather than with Wigner functions. Indeed, the time-independent Wigner functions usually obey\cite{Zachosbook} a homogeneous eigenvalue equation (such as wave functions), while the function $g({\bf R},\omega)$ is given by the inhomogeneous equation  \eqref{Dyson_star_1} of the Dyson type, which encloses in addition the causality principle.

Knowledge of the SO vortex Green's function allows one to compute 
quantum microscopic expressions of different observables. For that, the full 
Green's function, related to the SO vortex Green's function by Eq. 
\eqref{Green_function_nonlocal}, should be written in the electronic 
representation $\{|\mathbf{r} \rangle \}$. The latter Green's function is a $2 \times 2$ matrix 
in spin space  given by
${G}(\mathbf{r},\mathbf{r'},\omega)=\langle\mathbf{r}|\hat{G}(\omega)|\mathbf{r'}\rangle$, whose matrix elements will be written as $G_{\sigma 
\sigma'}(\mathbf{r},\mathbf{r}',\omega)$ with $\sigma,\sigma' \in \{\pm\}$. This change of representation (from vortex to electronic states) can be easily accomplished via a change of basis 
\begin{equation} 
G(\mathbf{r},\mathbf{r}',\omega)=\sum_{\nu_1,\nu_2}G_{\nu_1;\nu_2}(\omega)\tilde{\Psi}_{\nu_2}^\dagger(\mathbf{r}')\tilde{\Psi}_{\nu_1}(\mathbf{r}),  
\end{equation} 
where $\tilde{\Psi}_\nu(\mathbf{r})=\langle\mathbf{r}| \nu \rangle$ are the SO vortex wave functions 
defined by Eq. \eqref{SO_vortex_states}. Then, following Ref. 
\onlinecite{ChampelFlorens2008}, we can perform a change of coordinates 
$(\mathbf{R}_1,\mathbf{R}_2)\rightarrow(\mathbf{R}_{12},\mathbf{R}_{\textnormal{rel}})$
with $\mathbf{R}_\textnormal{rel}=(\mathbf{R}_2-\mathbf{R}_1)/2$, along with a 
Taylor expansion of the integrand to compute analytically the integral over the relative 
coordinates, $\mathbf{R}_\textnormal{rel}$, so that the electronic Green's function is finally written as an integral over the vortex position $\mathbf{R}_{12}$ only. 
In addition, making an integration by parts so that the operator  \eqref{Gaussian_exp} acts onto the product of
vortex  wave functions rather than on the local SO vortex Green's function, we obtain an exact expression  relating the 
electronic Green's function to the  
solution of Eq. \eqref{Dyson_star_1}:

\begin{multline}\label{electronicGreen2} 
 G_{\sigma\sigma'}(\mathbf{r},\mathbf{r}',\omega)=\int\dfrac{d^2\mathbf{R}}{2 \pi l_B^2} \sum_{n_1,\lambda_1}\sum_{n_2,\lambda_2} f_{\sigma}(\theta_{n_1}^{\lambda_1})f_{\sigma'}(\theta_{n_2}^{\lambda_2})\\ \times  F_{n_{1\sigma},n_{2\sigma'}}(\mathbf{r},\mathbf{r}',\mathbf{R}){g}_{n_{1},\lambda_{1};n_{2},\lambda_{2}}(\mathbf{R},\omega), 
\end{multline} 
where we have defined the kernel function 
\begin{equation}\label{kernel_function_general} 
F_{n_1,n_2}(\mathbf{r},\mathbf{r}',\mathbf{R})\equiv T^{-1}_{\mathbf{R}}\left[\Psi_{n_{2},\mathbf{R}}^\ast(\mathbf{r}')\Psi_{n_{1},\mathbf{R}}(\mathbf{r})\right], 
\end{equation}  
with $\Psi_{n,\mathbf{R}}(\mathbf{r})$ the vortex functions given in Eq. \eqref{vortex} and $f_\sigma(\theta_n^\lambda)$ defined in Eq. \eqref{weight_f} with the angular parameters \eqref{angles}. 
Equation \eqref{electronicGreen2} is nothing but the quantum formulation of the decomposition of the electronic motion into a cyclotronic motion [encapsulated in the kernel function \eqref{kernel_function_general}] superposed with a guiding center (or vortex) motion characterized by the Green's function $g_{n_{1},\lambda_{1};n_{2},\lambda_{2}}(\mathbf{R},\omega)$, which remains to be determined.

\subsection{Reduced matrix elements of the potential} 

The reduced matrix elements of the generalized potential $u_{n_1,\lambda_1;n_2,\lambda_2}(\mathbf{R})$  can be written as a sum of the (reduced) matrix elements of the scalar potential and of the contribution to the Hamiltonian which includes the fluctuations of the Rashba SO coupling parameter  
\begin{equation}\label{generalized_reduced_potential_matrix_elements} 
u_{n_{1},\lambda_{1};n_{2},\lambda_{2}}(\mathbf{R})=v_{n_{1},\lambda_{1};n_{2},\lambda_{2}}(\mathbf{R})+\delta {\cal H}_{n_{1},\lambda_{1};n_{2},\lambda_{2}}(\mathbf{R}). 
\end{equation}
Quite generally, the reduced matrix elements of the scalar potential read as
\begin{multline}\label{potential_matrix_elements_1} 
  v_{n_{1},\lambda_{1};n_{2},\lambda_{2}}(\mathbf{R}) = \sin (\theta_{n_{1}}^{\lambda_{1}}) \sin (\theta_{n_{2}}^{\lambda_{2}}) v_{n_1-1;n_2-1}(\mathbf{R})\\ + \cos (\theta_{n_{1}}^{\lambda_{1}}) \cos (\theta_{n_{2}}^{\lambda_{2}}) v_{n_1;n_2}(\mathbf{R}), 
\end{multline} 
where 
\begin{align}\label{potential_matrix_elements_2a} 
  v_{n_1;n_2}(\mathbf{R})&=T^{-1}_{\mathbf{R}}\langle n_{1},\mathbf{R}|\hat{V}|n_{2}, \mathbf{R} \rangle,\\ 
&=\int d^{2} \bm{\eta} \, F_{n_1,n_2}(\bm{\eta},\bm{\eta},\mathbf{0})V(\bm{\eta}+\mathbf{R}) \label{potential_matrix_elements_2b}
\end{align}
plays the role of an effective scalar potential seen by the vortex. Physically, it simply corresponds to an averaging of the bare scalar potential $V({\bf r})$ over the cyclotronic motion. As a result, the effective potential turns out to be always smoother than the bare one. Expression \eqref{potential_matrix_elements_1} characterizes the additional dependence of the total effective potential on the quantum number $\lambda$ resulting from the SO coupling. 

In high magnetic fields, it appears judicious to write alternatively the effective potential
 $v_{n_1;n_2}(\mathbf{R})$ as a series in powers of the magnetic length\cite{ChampelFlorens2008} 
\begin{equation}\label{potential_matrix_elements_expansion} 
  v_{n_{1};n_{2}}(\mathbf{R})=\sum^{+\infty}_{j=0}\sum^{+\infty}_{k=0} \dfrac{ (-\Delta_\mathbf{R})^j}{j!} \left(\dfrac{l_B}{2}\right)^{2j+k} \!\! v^{(k)}_{n_1;n_2}(\mathbf{R}), 
\end{equation} 
with coefficients 
\begin{multline}\label{potential_matrix_elements_expansion_2} 
 v^{(k)}_{n_{1};n_{2}}(\mathbf{R})=\dfrac{2^{k/2}}{k!}\sum^{k}_{l=0} \begin{pmatrix} 
 k \\ 
 l  
\end{pmatrix} 
\dfrac{(n_{1}+l)!}{\sqrt{n_{1}! \, n_{2}!}}\, \delta_{n_{1}+l,n_{2}+k-l}\\ \times(\partial_{X}+i\partial_{Y})^l  (\partial_{X}-i\partial_{Y})^{k-l} V(\mathbf{R}). 
\end{multline} 
Substituting this expansion in Eq. \eqref{potential_matrix_elements_1}, we see that at leading order  the total effective scalar potential  in the limit $l_B \rightarrow 0$ is clearly diagonal both in the level index $n$ and the SO quantum number 
\begin{equation}\label{potentialexpansion0} 
  v_{n_{1},\lambda_{1};n_{2},\lambda_{2}}^{(0)}(\mathbf{R})=\delta_{n_{1},n_{2}}\delta_{\lambda_{1},\lambda_{2}}V(\mathbf{R}). 
\end{equation}
The next (subdominant) terms in $l_B$ will produce mixing between different $n$ and $\lambda$ quantum numbers. The primary effect of the potential energy is thus to lift the energy degeneracy with respect to the guiding center by keeping the level index $n$ as a good quantum number. Note that, even when processes mixing $n$ are negligible, interesting effects related to a mixing of the two projections of $\lambda$ 
nevertheless occur at quadratic order in $l_B$ [processes involving second-order derivatives of the potential $V({\bf R})$], providing a mechanism for a spatial dispersion of the energy spin splitting.  

Working analogously, the reduced matrix elements of the Hamiltonian contribution describing the Rashba spatial fluctuations read as   
\begin{eqnarray}\label{matrix_elements_Rashba} 
 \delta {\cal H}_{n_1,\lambda_1;n_2,\lambda_2}(\mathbf{R})=-\dfrac{\hbar \sqrt{n_1}}{\sqrt{2}l_B} \Bigg[ 
\sin(\theta_{n_1}^{\lambda_1}) \cos(\theta_{n_2}^{\lambda_2})  \delta\alpha_{n_1;n_2}(\mathbf{R})  \nonumber
 \\ +
\cos(\theta_{n_1}^{\lambda_1}) \sin(\theta_{n_2}^{\lambda_2}) \delta \alpha_{n_1-1;n_2-1}(\mathbf{R}) 
   \Bigg] + (1\leftrightarrow2),  \hspace*{1cm}
\end{eqnarray}
where the notation $(1\leftrightarrow2)$ means exchanging indices $1$ and $2$ in the former expression. Here,  the matrix elements of the fluctuating Rashba parameter are defined in the same way as in Eqs. \eqref{potential_matrix_elements_2a} and \eqref{potential_matrix_elements_2b}, with the potential energy operator $\hat{V}$ replaced by the Rashba fluctuations $\widehat{\delta \alpha}$: 
\begin{align} 
 \delta \alpha_{n_1;n_2}(\mathbf{R})&=T^{-1}_{\mathbf{R}}\langle n_{1},\mathbf{R}|\widehat{\delta \alpha}|n_{2}, \mathbf{R} \rangle,\\ 
&=\int d^{2} \bm{\eta} \, F_{n_1,n_2}(\bm{\eta},\bm{\eta},\mathbf{0}) \, \delta \alpha(\bm{\eta}+\mathbf{R}).   \label{deltaalpha}
\end{align} 
Again, the quantity $\delta \alpha_{n_1;n_2}(\mathbf{R})$ can be regarded as an effective Rashba parameter resulting from the averaging of the Rashba fluctuations over the cyclotronic motion. Expanding similarly Eq. \eqref{deltaalpha} in powers of $l_B$, we get at high magnetic fields the leading contribution 
\begin{equation}
\delta {\cal H}^{(0)}_{n_1,\lambda_1;n_2,\lambda_2}(\mathbf{R})=-\dfrac{\hbar\sqrt{n_1}}{\sqrt{2}l_B} \sin(\theta_{n_1}^{\lambda_1}+\theta_{n_2}^{\lambda_2})  \delta_{n_1,n_2}\, \delta \alpha(\mathbf{R}),
\end{equation} 
which is still diagonal in the level index $n$, but now  predominantly  induces a mixing in the SO quantum number $\lambda$.

\section{Energy Spectrum for Drift States in Weakly Curved Scalar Potentials}\label{Spectrum_sec} 
\subsection{High magnetic field regime} 

So far, we have made no approximation until here, the Dyson equation \eqref{Dyson_star_1} being valid for any value of the external magnetic fields. We now focus on the high magnetic field regime which is characterized by negligible mixing between the integers $n$. This regime corresponds to considering  $\omega_c \rightarrow
+\infty$ while keeping $l_B$ finite (this is formally equivalent to the limit 
$m^\ast \rightarrow 0$). Only the matrix elements of the vortex Green's function diagonal in the level index $n$ are then relevant, so that we can write
 
\begin{align} 
  g_{n_{1},\lambda_{1};n_{2},\lambda_{2}}(\mathbf{R},\omega)& \simeq g_{n_{1};\lambda_{1};\lambda_{2}}(\mathbf{R},\omega) \, \delta_{n_{1},n_{2}}.
\end{align} 
As a result, Dyson equation \eqref{Dyson_star_1} takes the simpler form 
\begin{multline} 
 (\omega-E_{n,\lambda_1}+i0^+)g_{n;\lambda_1;\lambda_2}(\mathbf{R})=\delta_{\lambda_1,\lambda_2}\\ +\sum_{\lambda_3} u_{n;\lambda_1;\lambda_3}(\mathbf{R}) \star g_{n;\lambda_3;\lambda_2}(\mathbf{R},\omega), \label{Dyson}
\end{multline} 
where $u_{n_{1};\lambda_{1};\lambda_{2}}(\mathbf{R})=u_{n_{1},\lambda_{1};n_{2},\lambda_{2}}(\mathbf{R}) \, \delta_{n_1,n_2}$ reads 
\begin{equation}\label{generalized_matrix_elements_highfield} 
 u_{n;\lambda_1;\lambda_2}(\mathbf{R})=v_{n;\lambda_{1};\lambda_{2}}(\mathbf{R}) -\dfrac{\hbar\sqrt{2n}}{l_B}\sin(\theta_n^{\lambda_1}+\theta_n^{\lambda_2})\overline{\delta\alpha}_n(\mathbf{R}). 
\end{equation}
The matrix elements of the scalar  potential take the form
\begin{multline}\label{modifiedmatrixelements_scalar_highfield} 
 v_{n;\lambda_1;\lambda_2}(\mathbf{R})= \sin(\theta_n^{\lambda_1})\sin(\theta_{n}^{\lambda_2}){v}_{n-1}(\mathbf{R})\\+\cos(\theta_{n}^{\lambda_1})\cos(\theta_{n}^{\lambda_2}){v}_{n}(\mathbf{R}), 
\end{multline}
where the effective potentials $v_n({\bf R})\equiv v_{n;n}({\bf R})$ can be straightforwardly inferred from Eq. \eqref{potential_matrix_elements_2b}. The term $\overline{\delta\alpha}_n(\mathbf{R})$ defines the average of the Rashba fluctuations for the spin-split $n$th level 
\begin{equation}
 \overline{\delta\alpha}_n(\mathbf{R})=\dfrac{1}{2} \left[\delta \alpha_{n}(\mathbf{R})+\delta \alpha_{n-1}(\mathbf{R}) \right], 
\end{equation}
with $\delta \alpha_n({\bf R})\equiv\delta \alpha_{n;n}({\bf R})$.
 Note that in the high magnetic field regime, expression 
\eqref{generalized_matrix_elements_highfield} is symmetrical in the SO 
quantum numbers, i.e.,
$u_{n;\lambda_{1};\lambda_{2}}(\mathbf{R})=u_{n;\lambda_{2};\lambda_{1}}(\mathbf{R})$,
and therefore so must be the SO vortex Green's function 
$g_{n;\lambda_{1};\lambda_{2}}(\mathbf{R})=g_{n;\lambda_{2};\lambda_{1}}(\mathbf{R})$.

As already alluded to in the Introduction, the projection within a single Landau level
expresses the ineffective energy exchange between the guiding center (or vortex) motion and the cyclotronic motion, which are characterized by very different timescales in high magnetic fields. In the quantum realm, this exchange is only possible via a change of the Landau level index  (classically, this would correspond to a deformation of the cyclotron orbit with a change of the cyclotron radius) which is prohibited by a large  cyclotron gap in the spinless 2DEG. 
Whenever the Landau level mixing becomes negligible, the electron (more properly, the vortex) motion reduces to quasi-1D ballistic dynamics. The consideration of a finite magnetic length in the high magnetic field regime allows one to account for quantum effects within the 1D vortex dynamics, such as  interference effects responsible for tunneling or potential energy quantization. 

In the presence of SO coupling, the situation becomes somehow more tricky due to the possibility to induce transitions between different energy levels via the spin (extra) degree of freedom. In fact, neglecting level mixing between different $n$ can be 
justified whenever scattering due to the effective potential $u_{n_1,\lambda_1;n_2,\lambda_2}(\mathbf{R})$ 
from one state (with a given SO projection) to another one 
(necessarily, with opposite SO projection) is energetically forbidden due to the separation between the energy levels. Within the $l_B$ expansion of the effective potentials, we have seen in the previous section that the coupling between energy levels comes out with potential gradients, so that a simple quantitative (smoothness) criterion to neglect their mixing is

\begin{equation}\label{high_magnetic_condition} 
 l_B |\bm{\nabla}_\mathbf{R}U(\mathbf{R})| \ll |E_{n_2,\lambda}-E_{n_1,-\lambda}|. 
\end{equation} 
As pointed out in Sec. \ref{spinvortex}, the eigenenergies $E_{n,\lambda}$ in the pure case
 are no more equidistant in the presence of both SO and Zeeman  interactions. The  energy spectrum (cf. Fig. \ref{fig2}) is even characterized by level crossings for specific values of the parameter $S$. Clearly, level mixing processes, albeit small when considering a potential smooth at the scale of $l_B$, can not be neglected for these special points of the parameter space. However, as long as we are not working in the close vicinity of these points, inequality \eqref{high_magnetic_condition} tells us that we can safely ignore the mixing between the energy levels.

\subsection{Quantum drift states} 
\label{Drift}

The main difficulty in solving  Eq. \eqref{Dyson} resides in its differential character featured by the $\star$-product. Remarkably, this infinite-order differential operator reduces exactly to a simple product for any 1D potential $u_{n;\lambda_1;\lambda_2}({\bf R})$. For an arbitrary 2D potential, we expect that replacing the $\star$-product by a multiplication  is a very good approximation provided that the equipotential lines are relatively straight. This drift-state approximation amounts to describing the potential energy as a locally flat landscape, ignoring its Gaussian curvature which involves second-order derivatives of the potential in orthogonal directions. 
We have shown in a previous work\cite{ChampelFlorens2009} that this approximation is quantitatively valid for the local (thermal) vortex Green's function as long as the
curvature energy, which is a very small energy scale for a smooth potential, is smaller than the temperature energy scale. Curvature effects play an important role  to lift the 
quantum degeneracy of the Landau levels only at very low temperatures and close to the critical points of the potential landscape where the drift velocity vanishes. 
Within the context of the quantum Hall effect, this picture of quantum drift states in weakly curved
equipotential lines of the disorder potential has  been originally developed by Trugman\cite{Trugman1983} in terms of approximate Landau wave functions. We hereby formulate a similar implementation in terms of vortex Green's functions, which has the great advantage of not relying on a peculiar parametrization of the equipotential lines which can become especially cumbersome for a disordered potential landscape.

In this work, we shall only consider the leading drift approximation, thus ignoring the curvature effects. 
By making the approximation $\star \simeq \times $, the system of linear partial differential equations transforms  into a system of linear coupled algebraic equations which can be exactly solved by $2 \times 2$ matrix inversion. For $n \geq 1$, the resulting SO vortex Green's function presents a double pole structure 
\begin{multline}\label{green1D}
g_{n;\lambda_1;\lambda_2}(\mathbf{R},\omega)=\prod_{\epsilon=\pm}  \dfrac{1}{\omega-\xi_{n,\epsilon}(\mathbf{R})+i0^+} \Big\{ [\omega-E_{n,\lambda_1}\\-u_{n;\lambda_1;\lambda_2}(\mathbf{R})]\delta_{\lambda_1,\lambda_2}+u_{n;-\lambda_1;-
\lambda_2}\delta_{-\lambda_1,\lambda_2}\Big\}.                                 \end{multline} 
After some algebra involving Eqs. \eqref{relationorthogonality} and \eqref{generalized_matrix_elements_highfield}, the eigenenergies  given by the poles (including the potential energy contributions) are written as
\begin{equation}\label{spectrum1D} 
 \xi_{n,\epsilon}(\mathbf{R})=\hbar \omega_c \left[n-\dfrac{\epsilon}{2} \sqrt{\Delta_n(\mathbf{R})} \right]+ \overline{v}_n(\mathbf{R}), 
\end{equation} 
where  $\epsilon=\pm$  is a new SO quantum number (redefined due to the mixing between $\lambda_1$ and $\lambda_2$)
and 
\begin{eqnarray}\label{Delta} 
 \Delta_n(\mathbf{R}) &=& [1-Z_{n,\textnormal{eff}}(\mathbf{R})]^2+ n S_{n,\textnormal{eff}}^2(\mathbf{R}), 
\\ \label{averagepotential} 
 \overline{v}_n(\mathbf{R})& = & \dfrac{1}{2} \left[v_{n}(\mathbf{R})+v_{n-1}(\mathbf{R}) \right]. 
\end{eqnarray}
For $n=0$, the SO vortex Green's function $g_{0}({\bf R},\omega)$ is characterized by a single pole. Provided that we define $v_{-1}({\bf R}) \equiv 0$ and $\delta \alpha_{-1}({\bf R}) \equiv 0$, and impose that only the projection $\epsilon=-$ is allowed for the lowest energy level (as for $\lambda$ originally), the previous expressions hold also  for $n=0$.

Not surprisingly, energy spectrum \eqref{spectrum1D} presents a structure similar to that of the clean spectrum \eqref{eigenenergies_0}. The smooth potential energy contributions give rise to dressed Zeeman and Rashba SO couplings, which depend on the level index $n$ as well as the vortex position ${\bf R}$
\begin{eqnarray}\label{Z_eff} 
 Z_{n,\textnormal{eff}}(\mathbf{R}) &\equiv & Z+\delta Z_{n}(\mathbf{R})= Z -\dfrac{[v_{n}(\mathbf{R})-v_{n-1}(\mathbf{R})] }{\hbar \omega_c} , \hspace*{0.5cm}
\\
 S_{n,\textnormal{eff}}(\mathbf{R})& \equiv& S+\delta S_n(\mathbf{R}) = \dfrac{2 \sqrt{2}}{\omega_c l_B} \Big[\langle \alpha(\mathbf{r})\rangle+  \overline{\delta\alpha}_n(\mathbf{R}) \Big] \label{S_eff} .
\end{eqnarray} 
Quite interestingly, the spatially fluctuating parts of the effective Zeeman and Rashba coupling parameters have different origins. As naturally expected, (smooth) spatial fluctuations of the Rashba coefficient $\alpha({\bf r})$ lead  to a spatial dispersion of the energy spin splitting. In addition, spatial fluctuations of the scalar potential drive also another mechanism of spatial dispersion of the splitting via  an effective Zeeman coupling [and also via the average effective potential $\overline{v}_n( {\bf R})$ since pairs of spin-split energy levels involve different Landau level indices] which stems from the spinorial structure of the SO vortex wave functions.

It is worth noting that energy spectrum \eqref{spectrum1D} is still characterized   by accidental level crossings. The reason for this is that we have considered a projected Hamiltonian whose classical limit is integrable. We expect that even tiny mixing between the energy levels induced by any smooth disorder potential will induce level repulsion and give rise to tiny anticrossings in high magnetic fields. A thorough analysis of these special points in the parameter space (as a function of the magnetic field or the vortex position) is postponed for future work.

\subsubsection{Simple case of  a parabolic 1D potential} 

Now, we aim at analyzing the general energy spectrum \eqref{spectrum1D} in the simple situation of a parabolic 1D potential $V(\mathbf{R})$ in the presence of uniform Rashba coupling. In this case, only the level mixing processes between different $n$ have been neglected, since the drift approximation becomes exact (no curvature effects). With Eq. \eqref{potential_matrix_elements_expansion}, we straightforwardly get an explicit expression for the effective scalar potential for $n \geq 0$
\begin{equation}\label{modified_matrix_elements} 
v_{n}(\mathbf{R})= V(\mathbf{R})+\dfrac{l^{2}_{B}}{2} \left(n+\dfrac{1}{2} \right)\Delta_{\mathbf{R}} V(\mathbf{R}), 
\end{equation} 
and so for the averaged effective potential $\overline{v}_{n}(\mathbf{R})$ defined in Eq. \eqref{averagepotential}. According to Eq. \eqref{Z_eff}, the second-order derivatives of the potential lead to a modification of the effective Zeeman coupling with respect to the free case by the (constant) quantity
\begin{eqnarray}
\delta Z_n({\bf R}) = - \frac{l_B^2 \Delta_{{\bf R}} V({\bf R})}{2 \hbar \omega_c} \label{deltaZ},
\end{eqnarray} 
for $n \geq 1$. 

\begin{figure} 
\centering 
\includegraphics[width=0.43\textwidth]{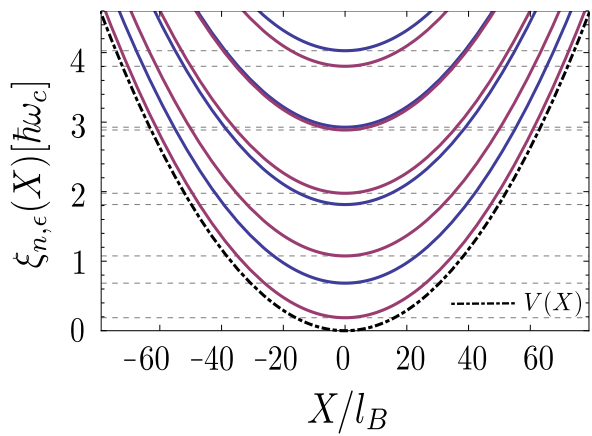} 
\caption{ (Color online)
Energy dispersion (in units of the cyclotron energy $\hbar \omega_c$) from Eq. \eqref{para}, as a 
function of the vortex position $X$, for the quadratic 1D potential  
(shown as a dashed-dotted parabola) in the absence of $n$ mixing.  
The characteristic length of the potential was chosen as $l_0=4 l_B$. 
The parameters are the same as in Fig.~\ref{fig2}, and 
the SO strength is $S_{\mathrm{exp}}=0.88$ as shown by the vertical line 
in Fig.~\ref{fig2}. The dashed horizontal lines underline the 
energies at $X=0$. 
Note that  
all the energy levels follow the same parabolic dispersion in this quadratic model. 
}
\label{fig3}  
\end{figure} 

As an illustration, let us take the 1D potential profile 
$V(x)=(1/2)m^\ast \omega_0^2 x^2$, with characteristic length scale 
$l_0=\sqrt{\hbar/(m^\ast \omega_0)}$. This profile describes the edge states toy model within the Hall bar geometry, for which no exact explicit solution valid at any magnetic fields is known analytically in the presence of Rashba coupling interaction. By neglecting mixing between different $n$, we get from the general result \eqref{spectrum1D} the following analytical expression for the eigenenergies:
\begin{eqnarray}
\xi_{n,\epsilon}(X) &=& n \hbar \omega_c \left[1+ \frac{1}{2} \left(\frac{\omega_0}{\omega_c}\right)^2 \right]+ \frac{\hbar \omega_0}{2 } \left( \frac{X}{l_0} \right)^2 \nonumber \\
&&-\frac{\epsilon}{2} \hbar \omega_c \sqrt{nS^2+ \left[1-Z+ \frac{1}{2} \left(\frac{\omega_0}{\omega_c}\right)^2\right]^2}. \hspace*{0.5cm} \label{para}
\end{eqnarray}
The energy dispersion thus consists in a set  of shifted parabolas as a function of the coordinate $X$  in the confinement direction, as shown in Fig. \ref{fig3}.  These parabolas, which appear in pairs with opposite quantum number, $\epsilon_1=-\epsilon_2$, present a uniform energy splitting,  because the effective Zeeman coupling \eqref{Z_eff} is independent of the position for globally quadratic potentials. We have checked that the obtained energy spin splitting between paired parabolas  is quantitatively consistent in high magnetic fields with the full numerical study of the quantum wire model performed in Ref. \onlinecite{Kramer2005}. The analytical solution \eqref{para} is closely related  to the energy spectrum derived in Ref. \onlinecite{Kushwaha2007} for the same simple quantum wire model, by using  the well-known mapping \cite{Kramer2005,Kushwaha2007,Emary2005}  in high magnetic fields of the quadratic electronic problem in the presence of Rashba and Zeeman interactions  to the 
exactly integrable Jaynes-Cummings Hamiltonian. However, the analytical approach developed in this paper to deal with the high magnetic field regime proves to be more general, since it allows one to also address the issue of (smooth) disorder effects.

\subsubsection{Interplay between spatial disorder and spin-orbit fluctuations in quantum drift states} 
\label{interplay_sec}  

The analysis of the quantum wire model reveals that a spatial dispersion of the energy spin splitting can not occur for a globally quadratic potential. A dispersion becomes nevertheless possible if the Laplacian of the potential varies spatially, as can be inferred from Eq. \eqref{deltaZ}.
In the case of an arbitrary smooth disorder potential, one has to consider the expression \eqref{spectrum1D} for the energy spectrum along with the general expressions \eqref{Z_eff} and \eqref{S_eff} for the effective Zeeman and SO coupling parameters.

\begin{figure}[t]
\centering 
\includegraphics[width=0.43\textwidth]{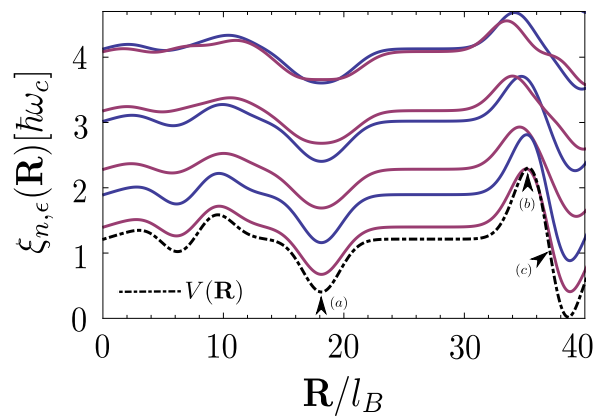} 
\caption{ (Color online) Energy dispersion (in units of the cyclotron energy $\hbar \omega_c$)  from Eq. \eqref{spectrum1D},
as a function of the vortex position ${\bf R}$, for a smooth random 1D disorder potential  (shown as a dashed-dotted line) in the absence of $n$ mixing. We use the same parameters as in Fig. \ref{fig2} with SO coupling strength $S_{\textnormal{exp}}=0.88$.
The energy  dispersion of the levels follows roughly the bare potential, but 
presents additional deviations, depending on the level $n$ and SO indices.  Weak level mixing processes (neglected here) between different $n$ will slightly lift the  degeneracies encountered at the observed crossings. The positions (a), (b), (c) correspond respectively to a local minimum, local maximum, 
and high gradient  region, and represent the three typical STM tip positions considered  in Fig.~\ref{fig5} when addressing the features of the local density of states.}
\label{fig4}  
\end{figure}

We illustrate the resulting energy dispersion in Fig. \ref{fig4} in 
the case of a random 1D potential $V({\bf r})$ (see the dashed-dotted line), assuming here a uniform Rashba coupling constant. Not surprisingly, all the energy levels follow roughly the bare potential, but more and more pronounced deviations can be seen for higher energy levels. A global trend is a flattening of the spatial dispersion when the level index $n$ increases. This simply results from stronger averaging effects with larger cyclotron radii in the average effective potential \eqref{averagepotential}.
 Furthermore, some energy levels, which were originally close in energy in the absence of the disorder potential, may cross at particular spatial positions. The unavoidable weak  mixing between $n$ levels (which is neglected here for simplicity) is expected to produce tiny anticrossings in place of the observed level crossings.
We can also notice small differences in the spatial dispersions of pairs of energy levels with opposite SO quantum number, which are direct consequences of the variations $\delta Z_{n}({\bf R})$ of the Zeeman effective coupling with position. The weakness of this effect is due to the relatively smooth spatial dependence of the effective potential $v_n({\bf R})$ at the scale of $l_B$ [see Eq. \eqref{Z_eff}].

As a result of the averaging over the cyclotron orbit, the spatial fluctuations $\delta S_{n}({\bf R})$ of the effective SO coupling parameter are also generally expected to be smooth. If the fluctuations are small in amplitude, it seems relevant to linearize the contributions  $\delta Z_{n}({\bf R})$ and $\delta S_{n}({\bf R})$ in the energy spectrum \eqref{spectrum1D}, so that
\begin{eqnarray}
\xi_{n,\epsilon}({\bf R}) \simeq E_{n,\epsilon}+\hbar \omega_c \frac{\epsilon}{2} \frac{(1-Z) \, \delta Z_{n}({\bf R})-nS \, \delta S_{n}({\bf R})}{\sqrt{(1-Z)^2+nS^2}} \nonumber \\
+\overline{v}_n({\bf R}). \hspace*{1cm}
\end{eqnarray}

Using the definition \eqref{Z_eff} of  $\delta Z_{n}({\bf R})$ in terms of the effective potential $v_n({\bf R})$, we can use the latter result to get a simple approximate expression for the energy spin splitting $ E_s({\bf R})=\xi_{0,-}({\bf R})-\xi_{1,+}({\bf R})$ within the first pair of energy levels [states $(1,+)$ and $(0,-)$]:
\begin{eqnarray}
E_s({\bf R})  \simeq  E_{0,-}-E_{1,+}
+\frac{ \hbar \omega_c}{2} \frac{ S \, \delta S_{1}({\bf R})}{\sqrt{(1-Z)^2+S^2}} \nonumber \hspace*{1.3cm} \\ 
+\frac{1}{2}\left[ 1-\frac{1-Z}{\sqrt{(1-Z)^2+S^2}}\right] \left[v_0({\bf R})-v_1({\bf R}) \right]. \hspace*{0.5cm} \label{Es}
\end{eqnarray}
In principle, both $\delta S_1({\bf R})$ and the difference in the effective potentials $\left[v_0({\bf R})-v_1({\bf R}) \right]$
can lead to a spatial dispersion of the spin splitting. Interestingly, the two mechanisms give rise to different dependencies on the magnetic field and on the level index $n$, which should help to discriminate between the two contributions to the energy in STS experiments. It is interesting to note that in the (classical) limit $l_B \to 0$, only the spatial contribution resulting from the fluctuations of the Rashba coupling coefficient remains. This indicates that the other dispersive contribution associated with the difference in effective potentials has a purely quantum-mechanical origin. These effects resulting from  the quantization of the cyclotron radius are notably more significant within the lowest  energy levels.

\section{Local Density of States}\label{LDoS_sec} 
\subsection{LDoS in weakly curved disorder potentials}  
 Before analyzing in more detail the experimental observations made in Refs. \onlinecite{Morgenstern2010,Morgenstern2012}, we aim at obtaining an analytical formula for the local density of states (LDoS) which, in addition to the energy spectrum, also contains information about the wave functions. Generally speaking, the spectral LDoS can be computed from the Green's function expressed in the electronic representation \eqref{electronicGreen2} and evaluated at coinciding electron positions, $\mathbf{r}=\mathbf{r}'$, by the general formula 
\begin{equation}\label{spectral_LDoS_definition} 
 \rho(\mathbf{r},\omega)=-\dfrac{1}{\pi} \textnormal{Im}\textnormal{Tr}\Big\{ G(\mathbf{r},\mathbf{r},\omega)\Big\}. 
\end{equation} 

In the high magnetic field regime ($\omega_c \rightarrow +\infty$ while $l_B$ finite), the vortex Green's function involved in Eq. \eqref{electronicGreen2} is diagonal in the level index $n$, so that only  the diagonal elements $F_{n,n}(\mathbf{r},\mathbf{r},\mathbf{R})\equiv F_{n}(\mathbf{r}-\mathbf{R})$ of the kernel functions \eqref{kernel_function_general} are needed. These form factors can be written as\cite{ChampelFlorens2009} 
\begin{align} 
F_{n}(\mathbf{r}-\mathbf{R})&= {\rm e}^{-(l_B^2/4)\Delta_\mathbf{R}}|\Psi_{n,\mathbf{R}}(\mathbf{r})|^2, \\
&=\dfrac{(-1)^n}{\pi l_B^2 } 
L_{n}\left[ \frac{2 ({\bf r}-{\bf R})^2}{l_B^2}\right] 
\, {\rm e}^{  -({\bf r}-{\bf R})^2/ l_B^2 }
\\ &=\dfrac{1}{\pi l_B^2 n!}\dfrac{\partial^n}{\partial s^n} \dfrac{{\rm e}^{-A_s (\mathbf{r}-\mathbf{R})^2/l_B^2 }}{1+s}\Bigg|_{s=0},\label{kernel} 
\end{align} 
with $A_s=(1-s)/(1+s)$ and $L_n(z)$ the Laguerre polynomial of degree $n$.
 The alternative writing \eqref{kernel} turns out to be specially convenient when considering the first levels. $F_{n}(\mathbf{r}-\mathbf{R})$ is an oscillating function that exhibits a sharp peak of width $l_B$ for  $|{\bf r}-{\bf R}|=R_n$, where $R_n=\sqrt{2n}l_B$ corresponds to the cyclotron radius.
Since the spinor weighting functions \eqref{weight_f} and the form factors  are purely real functions, the LDoS  can be
directly connected to the imaginary part of the SO
vortex Green's function:
\begin{equation}\label{LDoS_sum} 
 \rho(\mathbf{r},\omega)=\sum_{\sigma=\pm}\rho_\sigma(\mathbf{r},\omega), 
\end{equation} 
with the spin-projected LDoS given by
\begin{multline}\label{LDoS_per_spin} 
 \rho_\sigma(\mathbf{r},\omega)=-\dfrac{1}{\pi} \int \dfrac{d^2\mathbf{R}}{2 \pi l_B^2} \sum_{n=0}^{+\infty} \sum_{\lambda_1,\lambda_2} f_{\sigma}(\theta_{n}^{\lambda_1})f_{\sigma}(\theta_{n}^{\lambda_2}) \\ \times F_{n_\sigma}(\mathbf{r}-\mathbf{R}) \, \textnormal{Im}\, g_{n;\lambda_1;\lambda_2}(\mathbf{R},\omega). 
\end{multline} 
We remind here that $n_\sigma$ is defined in Eq. \eqref{nsigma} and have set $F_{-1}(\mathbf{r}-\mathbf{R})\equiv 0$ so that the above formula also holds for $n=0$.  

In a real STS experiment, the measured LDoS necessarily involves an extrinsic energy broadening caused by the temperature, which is taken into account by 
 a convolution of the spectral LDoS with the derivative of the Fermi-Dirac distribution. The LDoS per spin projection probed at energy $E$ is thus 
\begin{equation}\label{STS_LDoS_general} 
\rho_\sigma^{\textnormal{STS}}(\mathbf{r},E,T)=-\int d\omega \rho_\sigma(\mathbf{r},\omega)n'_{\textnormal{F}}(\omega), 
\end{equation} 
with 
\begin{equation}\label{derivative_FermiDirac} 
 n'_{\textnormal{F}}(\omega)=-\dfrac{1}{4 k_B T} \dfrac{1}{\cosh^2[(\omega-E)/2 k_B T]}.
\end{equation} 
The introduction of a finite temperature  fully justifies the recourse to a non-perturbative gradient expansion theory as developed in this work. The controlled character of the theory is granted in the vortex representation by the existence of a hierarchy of local energy scales formed by the successive spatial derivatives of the smooth effective potential generated by the $\star$-product differential operator [cf. Eq. \eqref{Dyson}]. The quantum drift approximation detailed in Sec. \ref{Drift} encapsulates the most robust quantum features associated with the (local) gradient of the potential, while a finite temperature allows one to disregard smaller (possibly inaccessible) energy scales characterizing more nonlocal quantum effects that can take place in a disordered potential landscape at zero temperature.
 
 Inserting the result \eqref{green1D} for the SO vortex Green's function established within the quantum drift approximation into previous formulas \eqref{LDoS_per_spin} and \eqref{STS_LDoS_general}, we obtain after  integration over the energy $\omega$ and  summation over $\lambda_1, \lambda_2$ an approximate analytical expression for the STS  LDoS per spin projection
\begin{multline}\label{LDoS} 
 \rho^{\textnormal{STS}}_\sigma(\mathbf{r},E,T)=\int \dfrac{d^2\mathbf{R}}{2 \pi l_B^2} \sum_{n=0}^{+\infty} \sum_{\epsilon=\pm} \left\{-n'_{\textnormal{F}}[\xi_{n,\epsilon}(\mathbf{R})]\right\} \\ \times F_{n_\sigma}({\bf r}-\mathbf{R})\dfrac{1}{2}\left[1+\epsilon \sigma\sqrt{1-\dfrac{nS_{n,\textnormal{eff}}^2(\mathbf{R})}{\Delta_n(\mathbf{R})}}\,\right].  
\end{multline} 
This formula constitutes  the main basis to interpret the recent STS experiments\cite{Morgenstern2010,Morgenstern2012}.

\subsection{Approximation for potentials smooth on the cyclotron radius} 

Clearly, even within the leading-order drift approximation, the electronic LDoS \eqref{LDoS} is a result of an intricate interplay between the drift and cyclotron degrees of freedom, described by the convolution of the form factor $F_{n_\sigma}({\bf r}-\mathbf{R})$ representing the circular motion with the (thermal) vortex spectral density. We can nevertheless get some useful analytical insight in particular limiting cases. 
An obvious simplification of Eq. \eqref{LDoS} occurs in the high-temperature (classical) regime, when $k_B T \gg R_n \left|{\bm \nabla}_{{\bf R}} \xi_{n,\epsilon}({\bf R}) \right|$. Under this inequality, we can essentially consider that ${\bf R} \simeq {\bf r}$ inside the functions depending smoothly on the vortex position, i.e., in the Fermi derivative function, as well as in the functions $S_{n,\mathrm{eff}}({\bf R})$ and $\Delta_n({\bf R})$. The only remaining dependence on the vortex position ${\bf R}$ contained in the kernel function is integrated out thanks to the normalization condition $\int d^2 {\bf R} \, F_{n}({\bf R})=1$. We then get the semiclassical expression for the total LDoS,
\begin{multline}
 \rho^{\textnormal{STS}}(\mathbf{r},E,T)= \dfrac{1}{2 \pi l_B^2} \sum_{n=0}^{+\infty} \sum_{\epsilon=\pm} \left\{-n'_{\textnormal{F}}[\xi_{n,\epsilon}(\mathbf{r})]\right\} ,  
\end{multline} 
which provides peaks of width $2 k_B T$ that are centered around the effective energies $\xi_{n,\epsilon}({\bf r})$ given by Eq. \eqref{spectrum1D}. Here, we remind that, according to the definition of the renormalized SO quantum numbers, only the projection $\epsilon=-$ is allowed for $n=0$.

In the opposite limit of small-temperature broadening, we can not entirely disregard the dependence of the eigenenergies on the vortex position ${\bf R}$. However, if the latter vary very smoothly on the scale of the cyclotron radius [$R_n$ is the typical characteristic lengthscale set by the kernel function $F_n({\bf R})$], it is reasonable to approximate this dependence up to the gradient contribution by writing 
  $\xi_{n,\epsilon}(\mathbf{R})\simeq \xi_{n,\epsilon}(\mathbf{r}) + \left(\mathbf{R}-{\bf r} \right) \cdot \nabla_\mathbf{r}\xi_{n,\epsilon}(\mathbf{r})$ [the other functions  $S_{n,\mathrm{eff}}({\bf R})$ and $\Delta_n({\bf R})$ are expanded similarly]. This linearization procedure is quite rough and the approximation will possibly break down if the electron starts to feel random fluctuations of the disorder potential on the scale of $R_n$.
Using the Fourier representation of the derivative of the Fermi-Dirac distribution \eqref{derivative_FermiDirac} 
\begin{equation} 
 n'_{\textnormal{F}}[\xi_{n,\epsilon}(\mathbf{R})]=- \int^{+\infty}_{-\infty} \dfrac{dt}{2 \pi} \dfrac{\pi Tt}{\sinh(\pi Tt)} {\rm e}^{it[\xi_{n,\epsilon}(\mathbf{R})-E]}, \label{nFourier}
\end{equation} 
and the expression for the kernels $F_{n}(\mathbf{R})$ given in Eq. \eqref{kernel}, we can perform the resulting Gaussian integration over the vortex position ${\bf R}$ in Eq. \eqref{LDoS} as it was detailed in Ref. \onlinecite{ChampelFlorens2010}. In the limit $T \rightarrow 0$, the integral over the variable $t$ in Eq. \eqref{nFourier} becomes then also purely Gaussian and can be straightforwardly evaluated. As a result, we get the following 
low-temperature expression for the spin-resolved LDoS: 
\begin{multline}\label{LDoS_lowT} 
 \rho_\sigma^{\textnormal{STS}}(\mathbf{r},E,0)\simeq \dfrac{1}{2 \pi l_B^2}\sum_{n=0}^{+\infty} \sum_{\epsilon=\pm}  \dfrac{1}{\sqrt{\pi} l_B \left|\nabla_\mathbf{r}\xi_{n,\epsilon}(\mathbf{r}) \right|} \dfrac{1}{2^{n_\sigma+1}}  \\ \times   \dfrac{1}{n_\sigma!} \left[1+\epsilon \sigma \sqrt{1-\dfrac{nS^2_{n,\textnormal{eff}}(\mathbf{r})}{\Delta_n(\mathbf{r})}} \,\right]H^2_{n_\sigma}\left[\dfrac{\xi_{n,\epsilon}(\mathbf{r})-E}{l_B \left|\nabla_\mathbf{r}\xi_{n,\epsilon}(\mathbf{r}) \right|}\right] \\ \times \exp \left\{-\left[\dfrac{\xi_{n,\epsilon}(\mathbf{r})-E}{l_B \left|\nabla_\mathbf{r}\xi_{n,\epsilon}(\mathbf{r}) \right|} \right]^2\right\}, 
\end{multline} 
where  $H_n(z)$ is the Hermite polynomial of degree $n$. A few comments are now in order. We first note the appearance of the local energy scale $l_B \left| \nabla_\mathbf{r}\xi_{n,\epsilon}(\mathbf{r})\right| $ associated with quantum drift. It gives rise to an intrinsic energy broadening of the LDoS peaks which can be roughly estimated as  $\sqrt{n} l_B \left| \nabla_\mathbf{r}\xi_{n,\epsilon}(\mathbf{r})\right| $ when including the spread resulting from the Hermite polynomials in addition to the Gaussian exponential factors. Expression
 \eqref{LDoS_lowT} is actually reminiscent  of the conventional LDoS formula for the (translation-invariant) Landau 
states generalized to the spinorial structure of the SO vortex wave functions. It presents additional asymmetries between the spin-up and -down states produced by the Rashba SO coupling. Finally, formula  \eqref{LDoS_lowT} obviously breaks down in the vicinity of the critical points where the drift energy $l_B \left| \nabla_\mathbf{r}\xi_{n,\epsilon}(\mathbf{r})\right| \simeq 0$. Close to these points, the finite temperature becomes again the main mechanism of LDoS broadening\cite{noteCurv}, and thus can no more be neglected. The different situations are all encompassed in the more general expression \eqref{LDoS}.

\subsection{Analysis of local spin splitting in STS experiments with InSb surface gases} 
\begin{figure} 
\centering
\includegraphics[width=0.43\textwidth]{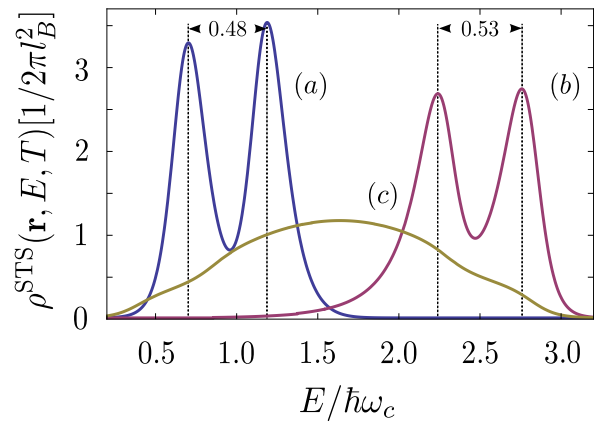} 
\caption{ (Color online)
Local density of states [in units of $1/(2\pi l_B^2)$] as a function of the  
energy $E$ for the three tip positions defined in Fig.~\ref{fig4}, 
focusing on the first two energy levels with the quantum numbers $(1,+)$ and $(0,-)$.  
The labels (a), (b), (c) correspond to the LDoS measured close to   minima, maxima and regions of strong gradient of the disorder potential, respectively (see 
Fig.~\ref{fig4}). The spin-split energy levels are only resolved in cases (a) and (b). The numbers on the top give the spin splitting in units of $\hbar \omega_c$, which in the present case are bigger   for maxima than minima of the potential landscape. This qualitatively reproduces the experimental features  reported in Refs.~\onlinecite{Morgenstern2010,Morgenstern2012}. An effective  temperature of $T = 15$ K was taken here to simulate additional broadening effects (temperature, experimental energy averaging, etc...).}
\label{fig5}  
\end{figure}

In the light of previous formulas, we are now in a position to interpret the recent  LDoS measurements \cite{Morgenstern2010,Morgenstern2012} performed at high magnetic fields in InSb surface gases. We show in Fig. \ref{fig5} the results of the calculation for the LDoS (focusing on the two lowest energy levels) on the basis of Eq. \eqref{LDoS} for the three different STM tip positions (a), (b) and (c) chosen in Fig. \ref{fig4} in the case of a 1D disordered potential and for a magnetic field of 7 Teslas. Of course, we have chosen here the same material parameters as in Figs. \ref{fig2} to \ref{fig4}, which are relevant for the experiments\cite{Morgenstern2010,Morgenstern2012}. We have also considered a uniform Rashba coupling, so that the spatial dispersion of the energies is only induced by the scalar electrostatic disorder.
The first observation is the impossibility to resolve the energy spin splitting in the case (c), corresponding to a STM tip position in a region of strong gradient of the disorder potential. In contrast, a spin-split LDoS can be clearly noticed for the two other tip positions (a) and (b) located in valley and hill regions of the potential landscape, respectively. These markedly different linewidths for the LDoS as a function of tip position, which were also found in experiments\cite{Morgenstern2010,Morgenstern2012}, can be understood if the peak broadenings are typically given by the drift energy scale  $l_B \left|{\bm \nabla}_{{\bf r}} \xi_{n,\epsilon}({\bf r}) \right| \propto l_B \left|{\bm \nabla}_{{\bf r}} V({\bf r}) \right|$, as expected in a low temperature regime (see previous section). Indeed, the drift energy gets strongly reduced 
  in potential hills or valleys due to the small drift velocity, while it may exceed the energy spin splitting in regions of strong potential gradients.

Another more subtle characteristic feature is the found correlation between the amplitude of the energy spin splitting and the disorder potential landscape, which is difficult to understand if the spatial variations of the spin splitting  are predominantly given by the local fluctuations of the Rashba SO coupling.
We have already established in Sec. \ref{interplay_sec} a mechanism induced by the combination of a uniform SO interaction and the potential disorder, which gives rise to a spatial dispersion of the spin splitting even in the absence of Rashba coupling fluctuations.
As reported in experiments \cite{Morgenstern2010,Morgenstern2012}, an enhanced spin splitting is found in hill regions in comparison with valley regions, with a variation of the order of 10 \% in Fig. \ref{fig5}. 
In the case of weak spatial fluctuations of the effective potential, we have derived  the estimation \eqref{Es} for the spin splitting between the first energy levels, which can  be further simplified for a very smooth disorder potential by using expansion  \eqref{potentialexpansion0}. As a result, we can directly correlate the spatial variations $\delta E_s({\bf R})$ of this spin splitting $E_s$ with the bare disorder potential $V({\bf R})$ via the simple analytical relation
\begin{eqnarray}
\delta E_s({\bf R})  \simeq
-\frac{l_B^2}{4}\left[ 1-\frac{1-Z}{\sqrt{(1-Z)^2+S^2}}\right] \Delta_{{\bf R}} V({\bf R}). \hspace*{0.5cm}
\end{eqnarray}
This formula helps now to understand why a larger $E_s(\mathbf{R})$ is found at hills of the disorder potential: The quantity $\Delta_{{\bf R}} V({\bf R})$ is typically negative at potential maxima, thus leading to $\delta E_s({\bf R})  >0$, i.e., to an enhancement of the spin splitting. At potential minima, $\Delta_{{\bf R}} V({\bf R})$ acquires an opposite sign, so that $E_s({\bf R})$ is generally reduced in valley regions. Note that, according to the general spectrum \eqref{spectrum1D}, this spatial correlation of the spin splitting with the potential landscape may be different when considering higher spin-split energy levels (e.g., it may be reverse depending on the magnetic field). Further STS experiments are required to make a thorough comparison with theory, specially to allow a more quantitative statistical analysis of the spatial fluctuations.

\section{Conclusion} 
In this paper, we have extended to 2DEG with Rashba SO coupling and  Zeeman interaction a semicoherent Green's function formalism well suited to study smooth disorder effects in quantum Hall systems. This formalism is based on the so-called SO vortex states,  which are spinful eigenstates of the clean  Hamiltonian that incorporate the topological  properties of the quantum motion of the electron (circular path around a singular point). The representation of the electronic quantum dynamics in terms of these states leads to a natural decomposition of the global motion into a cyclotronic motion and a  vortex (or guiding center) drift motion. We have shown that, at high magnetic fields, the electronic dynamics can be viewed as a vortex dynamics in the presence of an effective scalar electrostatic potential and an effective Rashba interaction. Dyson's equation of motion for the vortex Green's function has been solved for locally flat potential landscapes (quantum drift approximation), thus providing non-perturbative controlled expressions for the energy spectrum in the presence of smooth disorder. We have also derived an analytical formula for the LDoS at high magnetic fields taking into account spatial fluctuations of the Rashba coupling parameter, which should be useful for a future quantitative comparison between theory and LDoS measurements. We have shown that the intricate interplay between a smooth disorder potential and Rashba interaction leads to specific characteristic features for the spin-split LDoS, some of them having been already reported in recent STS experiments in InSb surface gases. The present work opens the door to a future thorough comparison between theory and experiment, especially at a more quantitative level, aiming at a better characterization of SO coupling in semiconductors.

\begin{acknowledgments} 
We would like to thank stimulating discussions with E. Ya. Sherman and M. Morgenstern. D. H. P.  acknowledges financial support from the RTRA Nanosciences Foundation in Grenoble.

\end{acknowledgments}


\begin{thebibliography}{39}
\bibitem{footnote1} One has to be careful in the association between the spin quantum number, $\sigma=\pm$, and the SO quantum number, $\lambda=\pm$, since they can only be identified in the limit of vanishing SO coupling, $\alpha \rightarrow 0$. In this limit, the angles $\theta_n^\lambda\rightarrow (1+\lambda)\pi/4$ and we have $f_{\pm}(\theta_n^\pm)=1$ so that the components of the spinor eigenstates of the Hamiltonian \eqref{Hamiltonian_0} decouple and can be written as tensor products in the form $|n,\mathbf{R},\pm \rangle=|n_{\pm},\mathbf{R} \rangle | \pm \rangle$ where $|\pm \rangle$ are the eigenstates of the Pauli matrix $\sigma_z$, $\sigma_z |\pm \rangle =\pm |\pm\rangle$.
\bibitem{noteCurv}
Actually, the curvature of the disorder potential gives also rise  at the critical points to a nonzero energy scale responsible for an intrinsic broadening of the LDoS, but we assume here that the temperature broadening overtakes the curvature effects.
\end{thebibliography}
\end{document}